# Estimation of friction force in an oscillator model of atomic force microscope tip sliding on vibrating surface

E. V. Kazantseva[a], Y. Braiman[b], J. Barhen[b]

[a]*Department of Solid State Physics and Nanosystems, National Research Nuclear University MEPhI (Moscow Engineering Physics Institute), Kashirskoye sh., 31, Moscow, Russia, 115409*
[b]*Center for Engineering Science Advanced Research, Computer Science and Mathematics Division, Oak Ridge National Laboratory, Oak Ridge, Tennessee, 37831*

**Abstract**

We consider an oscillator model to describe qualitatively friction force for an atomic force microscope (AFM) tip driven on a surface described by periodic potential. It is shown that average value of the friction force could be controlled by application of external time-dependent periodic perturbation. Numerical simulation demonstrates significant drop or increase of friction depending on amplitude and frequency of perturbation. Two different oscillating regimes are observed, they determined by frequency and amplitude of perturbation. The first one is regime of mode locking at frequencies multiple to driving frequency. It occurs close to resonance of harmonic perturbation and driving frequencies. Another regime of motion for a driven oscillator is characterized by aperiodic oscillations. It was observed in the numerical experiment for perturbations with large amplitudes and frequencies far from oscillator eigenfrequency. In this regime the oscillator does not follow external driving force, but rather oscillates at several modes which result from interaction of oscillator eigenmode and perturbation frequency.
PACS: 05.45.Tp, 62.20.Qp, 62.25.Fg, 68.37.Ps

## I. INTRODUCTION

Several theoretical works describe phenomenon of friction at nanoscale level. Experimental and numerical investigation of stick-slip processes [1] is performed in [2]. The dynamics of an array of the coupled particles driven in periodic potential in presence of external force is analyzed [3]. This theoretical model enhanced with additional effect of fluctuations caused by Gaussian noise was considered in [4]. One-particle Tomlinson-type model is considered in Ref. [5] and the low-friction regime was observed for certain amplitudes and frequencies. The difference between friction reduction mechanism in stick-slip and sliding regimes is discussed. In [6] drop in friction coefficient for certain range of frequencies was observed in experiment with AFM in study of effect of normal vibrations on friction coefficient. It was pointed out that vibrations of periodic surface may result in significant reduction of friction; a mechanism of friction reduction is proposed based on numerical simulation of the simplified model. Stick-slip motion was observed in numerical modeling [7] of coupled oscillators (oscillator array) in a periodic potential subjected to constant external force and noise. The model of the AFM tip represented by an array of coupled oscillators subjected to external force and driven over surface with periodic lattice potential is considered in [8]. The regimes of oscillations with very small and negative friction were found for a certain range of perturbation frequencies.

Oscillator model was implemented in [9] to provide theoretical explanation of the experiment on measuring friction of sliding alumina pin on steel sample with or without debris blown by air jet. The authors revealed steady regime with constant friction force and stick-slip regime characterized either periodic or irregular fluctuations of friction force. Time series and power spectrum obtained in experiment were compared with results of numerical simulation of oscillator model with several parameters.

Forced oscillator model is effectively used to describe another physical system, one or several coupled Josephson junctions with imposed bias current [10-12]. In oscillator model of Josephson junction a variable for an oscillator displacement corresponds to phase difference of wave function of current density at junction.

In theoretical investigation we study how the friction force could be controlled by external time-dependent perturbation and driving. Using numerical simulations and analytical approximations we found conditions beneficial for realization of different regimes of AFM tip motion. In driven oscillator model with perturbation the force of friction depends on character of oscillatory regime. We consider one-oscillator model and perform numerical simulation to reveal how the friction force can be controlled by variation of amplitude and frequency of external perturbation. The time series and frequency spectra of the friction force are analyzed. Two different types of spectrum are observed: a discrete set of frequency components (mode-locked regime), and broad continuous spectrum, corresponding to many frequencies generated in nonlinear parametric process. For mode-locked regime the friction force can be represented as a set of harmonics and subharmonics which are proportional to driving frequency $v$. In particular cases, when the major input to the spectrum of the friction force comes basically from a small number of harmonics, we propose an approximation for the friction force and compare it with its time series. Average value of friction force is estimated from this approximation.

The paper is organized as follows. In Sec. 2 we consider a model equation used to simulate oscillations of an AFM tip which slides over surface with periodic lattice potential and subjected to time-dependent perturbation. In Sec. 3 we present the results of numerical simulation of model equation. The average value of friction force is calculated over time interval of integration, these results are presented as dependence of friction force on frequency and amplitude of perturbation. Two different regimes of oscillation are observed in numerical simulations. One type of AFM tip movement over a surface in presence of periodic perturbation is a mode-locked regime. It is characterized by quasiperiodic time series of friction force and discrete frequency spectrum. Another regime of motion is characterized by chaotic oscillations of friction force, whose frequency spectrum is continuous. In Sec. 4 two different types of oscillator behavior (mode-locked and aperiodic) are discussed. In Sec. 5 the results are summarized.

## II. MODEL EQUATION

A theoretical model of AFM tip displacement is described by motion equation for oscillator subjected to viscous friction and elastic forces, driving of AFM tip with constant velocity and sliding over surface described by periodic lattice potential $U(x) = -U_0 \cos x$.

$$m\frac{d^2x}{dt^2} = F_{vf} + F_e + F_s + F(x). \qquad (1)$$

Viscous friction force $F_{vf} = -\eta l \frac{dx}{dt}$, $\eta$ – viscosity, $l$ – length parameter, elastic force $F_e = -kx$, $k$ – stiffness or spring constant, $x$ – displacement. Driving force $F_s = m\omega_s^2 Vt$, with frequency $\omega_s$, velocity $V$. Restoring force determined by interaction with lattice potential $F(x) = -\partial U/\partial x = F_0 \sin x$.

$$m\frac{d^2x}{dt^2} + \eta l \frac{dx}{dt} + kx - m\omega_s^2 vt + F_0 \sin x = 0. \qquad (2)$$

In this equation variable $x$ defines displacement of oscillator which models the AFM tip. Oscillator eigenfrequency $\omega_0 = \sqrt{k/m}$. The effect of periodic perturbation normal to sliding surface on dynamics of AFM tip is considered as an additional term included into model equation. It is characterized by amplitude $A$ and frequency $\Omega$.

$$\frac{1}{\omega_0^2}\frac{d^2x}{dt^2} + \frac{\eta l}{m\omega_0^2}\frac{dx}{dt} + x - \frac{\omega_s^2}{\omega_0^2}Vt + \frac{F_0}{m\omega_0^2}(1 + A\sin\Omega t)\sin x = 0. \qquad (3)$$

Introduce new evolution variable $\varphi = \omega_0 t$, then Eq. (3) results in following:

$$\frac{d^2x}{d\varphi^2} + \frac{\eta l}{m\omega_0}\frac{dx}{d\varphi} + x - \frac{\omega_s^2}{\omega_0^2}\frac{V}{\omega_0}\varphi + \frac{F_0}{m\omega_0^2}\left(1 + A\sin\left(\frac{\Omega}{\omega_0}\varphi\right)\right)\sin x = 0. \qquad (4)$$

To simplify presentation of theoretical and numerical results a dimensionless evolution variable $t$ is used in following considerations instead of phase variable $\varphi$. There are new parameters defined, they are $\gamma = \eta l/(m\omega_0)$, $v = V/\omega_0$, $\omega = \Omega/\omega_0$ and $\omega_s = \omega_0$, $x_1 = F_0/(m\omega_0^2)$. Parameter $v$ and displacement $x$ have physical dimension of length. Decay constant $\gamma$, perturbation frequency $\omega$ and time variable $t$ are dimensionless. Then the Eq. (4) could be written as

$$\frac{d^2x}{dt^2} + \gamma\frac{dx}{dt} + x - vt + x_1(1 + A\sin\omega t)\sin x = 0. \qquad (5)$$

This oscillator equation describes dynamics of the AFM tip. The Eq. (5) is similar to Eq. (3) of Ref. [8]. However, in that work a dimensionless oscillator displacement was considered as the phase similarly to phase difference $\theta$ of the wave function of current density at the contacts in theoretical model of oscillator describing Josephson junction [10].

The average friction force is defined as

$$F_{frict} = \frac{k_1}{T}\int_0^T (vt - x(t))dt. \qquad (6)$$

$T$ is the time interval over which averaging (integration) takes place. Friction coefficient is set to unity $k_1=1$, and average friction force determines average detuning of oscillator coordinate from driving parameter $v$ along observation period. The normalization parameter $x_1$ of oscillator displacement due to interaction with substrate lattice in Eq. (5) is also set to unity, $x_1 = 1$, in further numerical and theoretical results.

Numerical solution of the oscillator equation Eq. (5) and evaluation of average friction force with the Eq. (6) is performed in this work. It occurs that the friction force drops or increases at the perturbation frequencies proportional to multiples of the driving parameter $v$. Characteristic resonant frequencies are $nv$, $\omega_0 \pm lv$ (with $n, l$ - integers). The amplitude of average friction force attains for its extreme at perturbation frequencies proportional to driving parameter $v$. Global maximum is observed at resonance of driving parameter $v$ and frequency of perturbation $\omega = v$ at $\gamma = 0.1 \div 0.2$. For the same range of decay constant a local minimum of average friction force is attained for certain amplitudes of perturbation with frequency twice as value of driving parameter: $\omega = 2v$. Also negative value of average friction force is obtained in numerical simulations for intervals of perturbation amplitudes. In numerical simulation described in this work the initial values of coordinate and velocity are set at equilibrium $x = \dot{x} = 0$. Decay constant $\gamma = 0.05 \div 0.2$ was considered in numerical simulations. Forced AFM tip sliding is characterized by parameter $v = 0.1$. This value has been used in numerical calculations of friction force in a similar oscillator model [6, 8].

### III. NUMERICAL SIMULATION

#### A. Dependence of friction force on frequency and amplitude of perturbation in numerical simulation

The Fig. 1 illustrates the influence of the perturbation amplitude at frequency characteristics of the friction force.

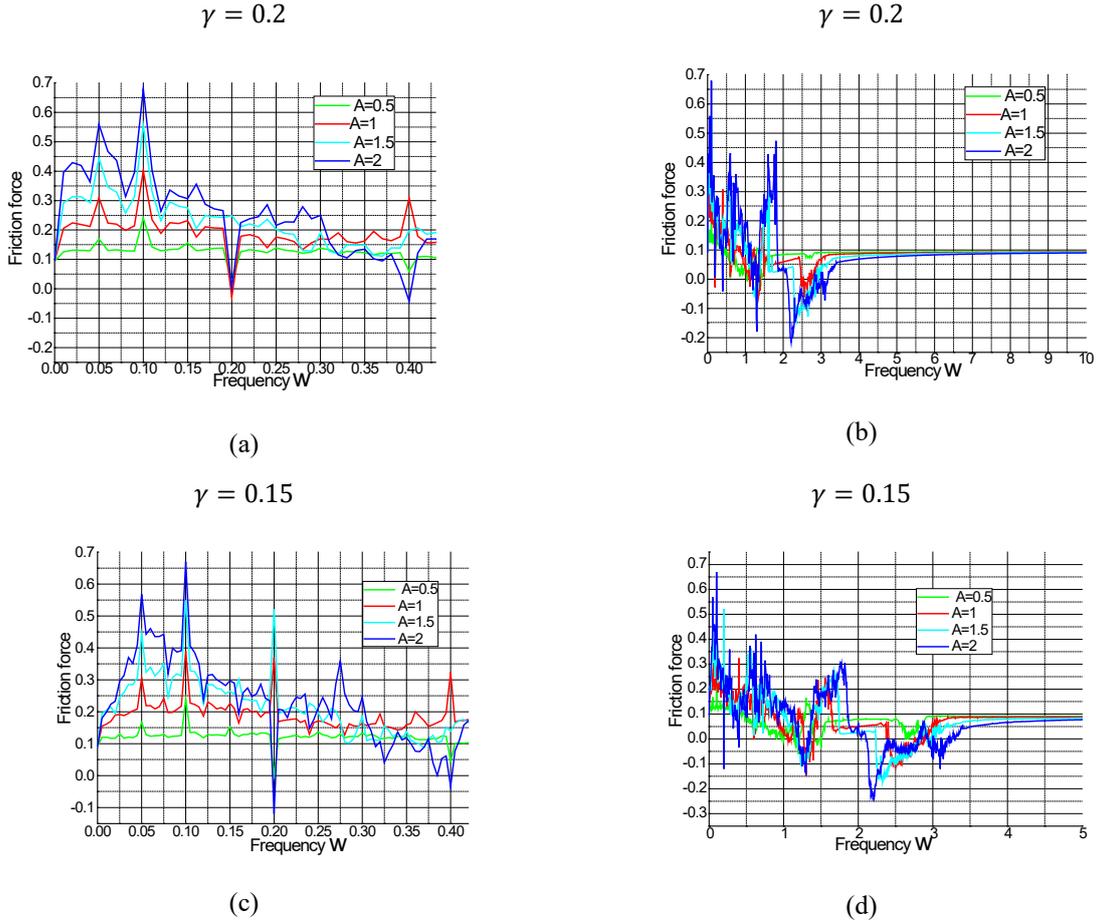

FIG. 1. The plots of average friction force in dependence on the frequency $\omega$ of perturbation with amplitude $A$. Decay constant $\gamma = 0.2$ for panels (a)-(b) and $\gamma = 0.15$ for panels (c)-(d). The green solid line corresponds to $A = 0.5$, red is for $A = 1$, cyan is for $A = 1.5$ and blue solid line corresponds to $A = 2$.

In the Fig. 1 are shown the plots of average friction force as a function of perturbation frequency for values of decay constant $\gamma = 0.2$ at the Fig. 1(a)-(b) and $\gamma = 0.15$ at the Fig. 1(c)-(d). The force of friction depends considerably on the amplitude of perturbation with certain frequency until $\omega > 4$, as shown at the Fig. 1(b), (d), where it tends to constant value corresponding to unperturbed oscillator, at $\omega = 0$ perturbation term $A\sin\omega t = 0$, as shown at the Fig. 1(a), (c). Friction force decreases when perturbation frequency tends to oscillator's eigenfrequency: $\omega \to \omega_0 = 1$. In perturbation's frequency range $1 < \omega < 3$ friction force becomes very low or negative for certain amplitudes and frequencies of perturbation. For all amplitudes $A = 0.5, 1, 1.5, 2$ the friction force attains for local maximum at perturbation frequency $\omega = \nu/2 = 0.05$ and global maximum at $\omega = \nu = 0.1$, as illustrated by Fig. 1(a), (c). At $\omega = 2\nu = 0.2$ a local minimum is observed at the Fig. 1(a) for perturbation amplitudes $A = 0.5, 1, 2$ and decay constant $\gamma = 0.2$ and at the Fig. 1(c) for $A = 0.5, 2$ when decay constant $\gamma = 0.15$; a local maximum is observed at $A = 1.5, \gamma = 0.2$ and $A = 1, 1.5, \gamma = 0.15$. If perturbation frequency $\omega = 4\nu = 0.4$ a local maximum is observed at $A = 1$, a local minimum appears for perturbation amplitudes $A = 0.5, 2$ both for $\gamma = 0.2$ and $\gamma = 0.15$, as shown at the Fig. 1(a), (c).

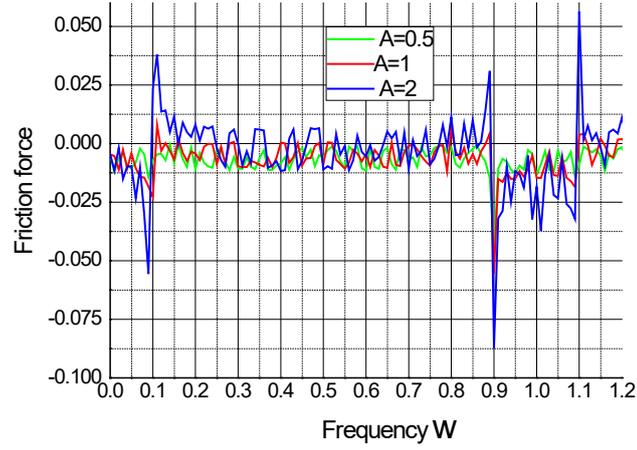

(a)

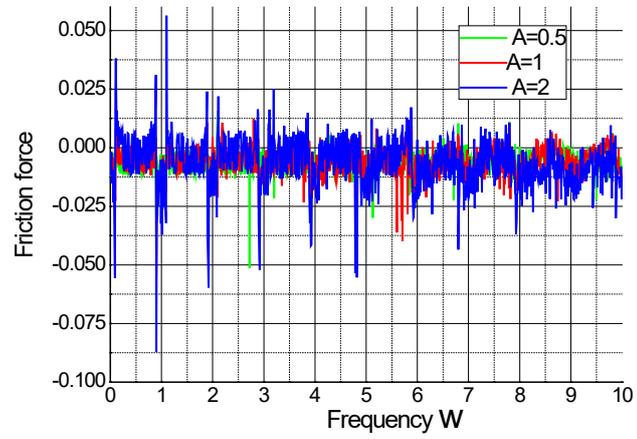

(b)

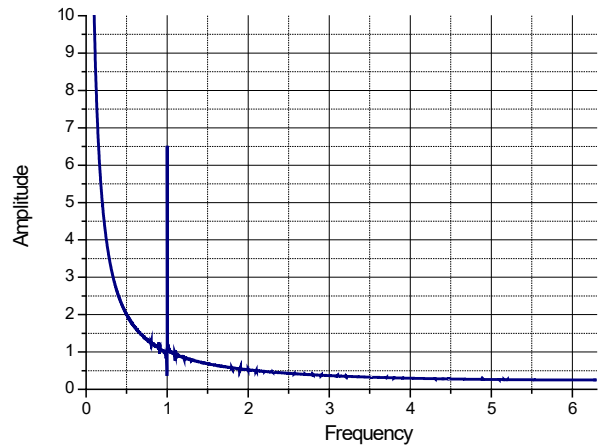

(c)

FIG. 2. (a)-(b) The plots of average friction force in dependence on the frequency $\omega$ of perturbation with amplitude $A$. Decay constant $\gamma = 0.05$. The green solid line corresponds to $A = 0.5$, red is for $A = 1$, and blue solid line corresponds to $A = 2$. (c) Frequency spectrum of the friction force $f_{frict}(t) = vt - x(t)$ for perturbation with $A = 1$, $\omega = v = 0.1$.

If decay constant $\gamma = 0.05$ average friction force is low at the whole interval of perturbation frequencies, varying from positive to negative values at different frequencies. It shows resonant behavior when perturbation frequency $\omega = v$, and $\omega = \omega_0 \pm v$, as seen at the Fig. 2(a). For perturbation with frequency $\omega = v = 0.1$ and amplitude $A = 1$, there is a single mode with spectral amplitude $\alpha_{\omega=1} = 6.5$ in frequency spectrum shown in Fig. 2(c). Spectral amplitude $\alpha_{\omega=1} = 4.2$ for $A = 0.5$ and $\alpha_{\omega=1} = 7.7$ when $A = 2$. This discrete mode corresponds to wave generation at oscillator's eigenfrequency $\omega_0$, continuous spectrum appears because oscillator's amplitude grows with time. Integration interval of time is [0:5000]. Friction force time series at $t > 500$ are approximated as

$$f_{frict}(t) = vt - x(t) = \sin(t + \varphi_0) * t^{a+b*t}. \quad (7)$$

At $t < 500$ carrier wave of friction force is the same function $\sin(t + \varphi_0)$, but it's envelope below approximating function $t^{a+b*t}$. When perturbation amplitude $A = 0.5, \varphi_0 = 2.3, a = 0.29, b = 8 * 10^{-6}$; if perturbation amplitude $A = 1, \varphi_0 = 1, a = 0.3, b = 9 * 10^{-6}$; if perturbation amplitude $A = 2, \varphi_0 = -0.5, a = 0.35, b = 7 * 10^{-6}$.

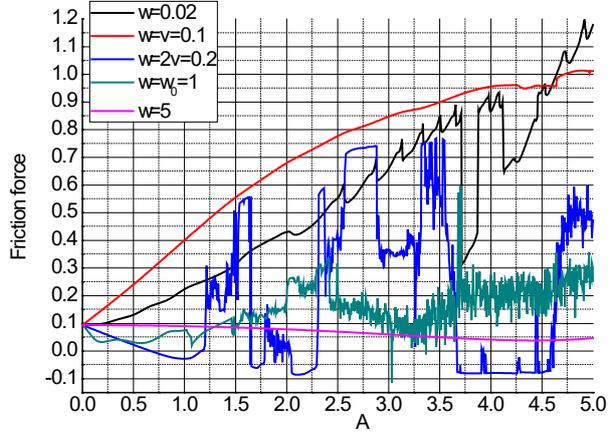

(a)

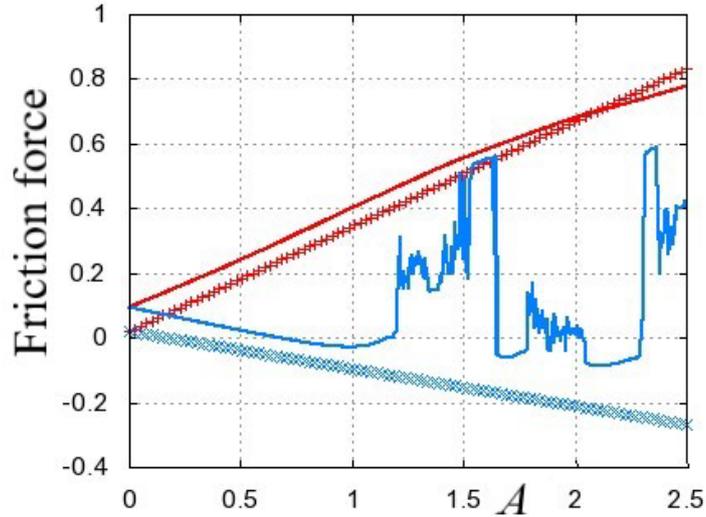

(b)

FIG. 3. Dependence of average friction force on amplitude $A$ of perturbation with frequency $\omega$. Decay constant $\gamma = 0.2$. (a) Black solid line corresponds to $\omega = 0.02$, red line corresponds to $\omega = v = 0.1$, blue line to $\omega = 2v = 0.2$, blue-green line to $\omega = 1$, magenta solid line corresponds to $\omega = 5$. (b) $F_{frict}(A)$ at $\omega = v$ (red solid line) and $\omega = 2v$ (blue solid line) and their approximate graphs at $\omega = v$ (red points) and $\omega = 2v$ (blue points).

As can be seen from the Fig. 3(a), for perturbation frequency $\omega = \nu$ (red curve) average friction force is approximately a linear function provided that perturbation amplitude $A<2.5$. For perturbation frequency $\omega = 2\nu$ (blue curve) average friction force decreases with $A$ up to a value of $A = 1$; in the amplitude region $1 < A < 5$ friction force close to constant at several frequency intervals and it sharply changes from positive to negative values between these intervals. At very low perturbation frequency $\omega = 0.02$ (black curve) friction force smoothly increases up to $A \approx 2.5$ and at $A > 2.5$ average friction force $F_{frict}(A)$ grows with perturbation amplitude as a sawtooth function. At perturbation frequency $\omega = 1$ (blue-green curve) average friction force smoothly decreases with increase of perturbation amplitude up to $A \approx 0.7$, increases up to $A \approx 2$ and fluctuates at $A > 2.2$. At high frequencies ($\omega > 4$) of perturbation friction force almost does not change with increase of perturbation amplitude, at Fig. 3(a) magenta curve corresponds to $\omega = 5$.

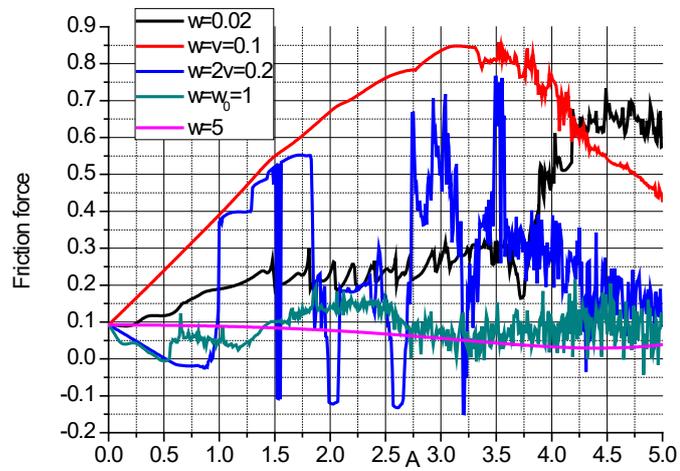

(a)

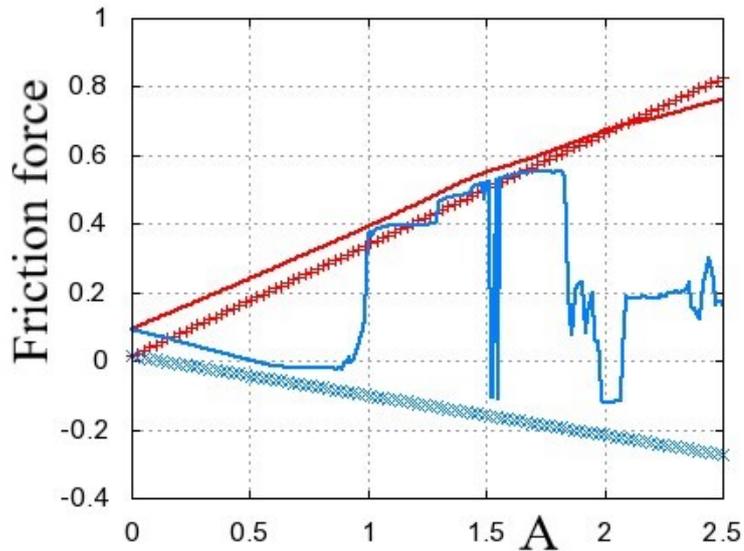

(b)

FIG. 4. Dependence of average friction force on amplitude $A$ of perturbation with frequency $\omega$. Decay constant $\gamma = 0.15$. (a) Black solid line corresponds to $\omega = 0.02$, red to $\omega = \nu = 0.1$, blue to $\omega = 2\nu = 0.2$, blue-green to $\omega = 1$, magenta solid line corresponds to $\omega = 5$. (b) $F_{frict}(A)$ at $\omega = \nu$ (red solid line) and $\omega = 2\nu$ (blue solid line) and their approximate graphs at $\omega = \nu$ (red points) and $\omega = 2\nu$ (blue points).

As shown in the Fig. 4(a) for perturbation frequency $\omega = \nu$ (red curve) average friction force increases up to perturbation amplitude $A \approx 3.25$ if $\gamma = 0.15$. For perturbation frequency $\omega = 2\nu$ (blue curve) friction force decreases with $A$ if $A < 0.9$, and in the amplitude region $1 < A < 5$ friction force sharply changes from positive to negative values between frequency interval of constant friction force.

In the Fig. 3(b) and Fig. 4(b) average friction force dependence on amplitude $A$ of perturbation with frequencies $\omega = \nu$ (solid red curve) and $\omega = 2\nu$ (blue solid curve) is compared with results of one-mode approximation

$$x(t) = \nu t - \alpha_1 \sin(\nu t), \qquad (8)$$

used in Appendix A to estimate friction force. Approximate graphs at perturbation frequencies $\omega = \nu$ (red points) and $\omega = 2\nu$ (blue points) are plotted with points using the same colors as for numerically obtained data. Estimation of friction force in one mode approximation for perturbation with amplitude $A$ and frequency $\omega = \nu$ results in following (see Eq. (A11) in Appendix A)

$$F_{frict}(\omega = \nu, A, \alpha_1 = 1) = \gamma \nu + A\big(J_0(1_1) - J_1(1)\big). \qquad (9)$$

Parameters are $\alpha_1 = 1, \nu = 0.1$. If decay constant $\gamma = 0.2$ for different perturbation amplitudes $A$ friction force is following: $F_{frict}(A = 0.5) \approx 0.02 + 0.16 = 0.18$, $F_{frict}(A = 1) \approx 0.02 + 0.32 = 0.34$, $F_{frict}(A = 2) \approx 0.02 + 0.65 = 0.67$. At $\gamma = 0.15$ $F_{frict}(A = 0.5) \approx 0.015 + 0.16 = 0.165$, $F_{frict}(A = 1) \approx 0.335$, $F_{frict}(A = 2) \approx 0.67$ (as shown in the system of Eqs. (A12) in Appendix A).

Linear approximation of function $F_{frict}(A)$ by Eq. (9) provides at $A < 2$ same slope coefficient as in numerically obtained data. Constant term $\gamma \nu$ obtained in one-mode approximation is less than average friction force $F_{frict}(A = 0)$ in numerical results for oscillator without perturbation.

If perturbation frequency $\omega = 2\nu$ friction force in the same approximation is (see the Eq. (A14) in Appendix A)

$$F_{frict}(\omega = 2\nu, A, \alpha_1 = 1) = \gamma \nu - \frac{A}{2}\big(J_1(1) + J_3(1)\big) = \gamma \nu + A\big(J_0(1) - 2J_1(1)\big). \qquad (10)$$

For different perturbation amplitudes $A$ at $\gamma = 0.2$ $F_{frict}(\omega = 2\nu, A = 0.5) \approx -0.037$, $F_{frict}(\omega = 2\nu, A = 1) \approx -0.095$, $F_{frict}(\omega = 2\nu, A = 2) \approx -0.21$. When $\gamma = 0.15$ $F_{frict}(\omega = 2\nu, A = 0.5) \approx -0.042$, $F_{frict}(\omega = 2\nu, A = 1) \approx -0.1$, $F_{frict}(\omega = 2\nu, A = 2) \approx -0.215$, as shown in Appendix A in the system of Eqs. (A15).

For perturbation frequency $\omega = 2\nu = 0.2$ linear approximation of function $F_{frict}(A)$ by Eq. (10) predicts negative values of average friction force if $A\big(J_1(\alpha_1) + J_3(\alpha_1)\big) > 2\gamma \nu$ and provides up to $A \approx 1$ same slope coefficient as in numerically obtained data. Eq. (10) does not explain spectral intervals with constant positive average friction force which appear at higher amplitudes of perturbation.

When perturbation occurs at high frequency $\omega = n\nu, n \gg 1$,

$$F_{frict} = \gamma \nu + \frac{\nu A}{4\pi}\int_0^{2\pi/\nu}\big(\cos(n\nu t + \alpha_1 \sin(\nu t)) - \cos(n\nu t - \alpha_1 \sin(\nu t))\big)dt \approx \gamma \nu. \qquad (11)$$

Average friction force is small and does not depend on amplitude of perturbation with frequency $\omega > 4$. It could be explained because oscillator would not respond to external force with frequency beyond its eigenfrequency. In subsection C is considered two-mode approximation which provides numerical value of $F_{frict}(A = 0)$ (without perturbation) close to average friction force calculated in numerical simulation. It seems plausible that there are different mechanisms contributing into friction force at different perturbation frequencies if to compare the curves of friction force illustrated at the Figs. 1-4.

## B. Time series of oscillator in presence of perturbation

From definition of average friction force in the Eq. (6) it is follows that the time series of the coordinate $x(t)$ are required to calculate this integral. If oscillatory process is periodic in time and $x(t)$ is a periodic function, the average over whole time interval can be substituted by average over this period. For quasiperiodic process the time series of $x(t)$ are obtained from numerical experiment and estimated with a set of periodic functions. This approximation is used to calculate average friction force. As an example the time series are shown in the Fig. 5 for the friction force $f_{frict} = vt - x(t)$ at driving parameter $v = 0.1$ for various values of amplitude and frequency of perturbation. Decay constant $\gamma = 0.15$.

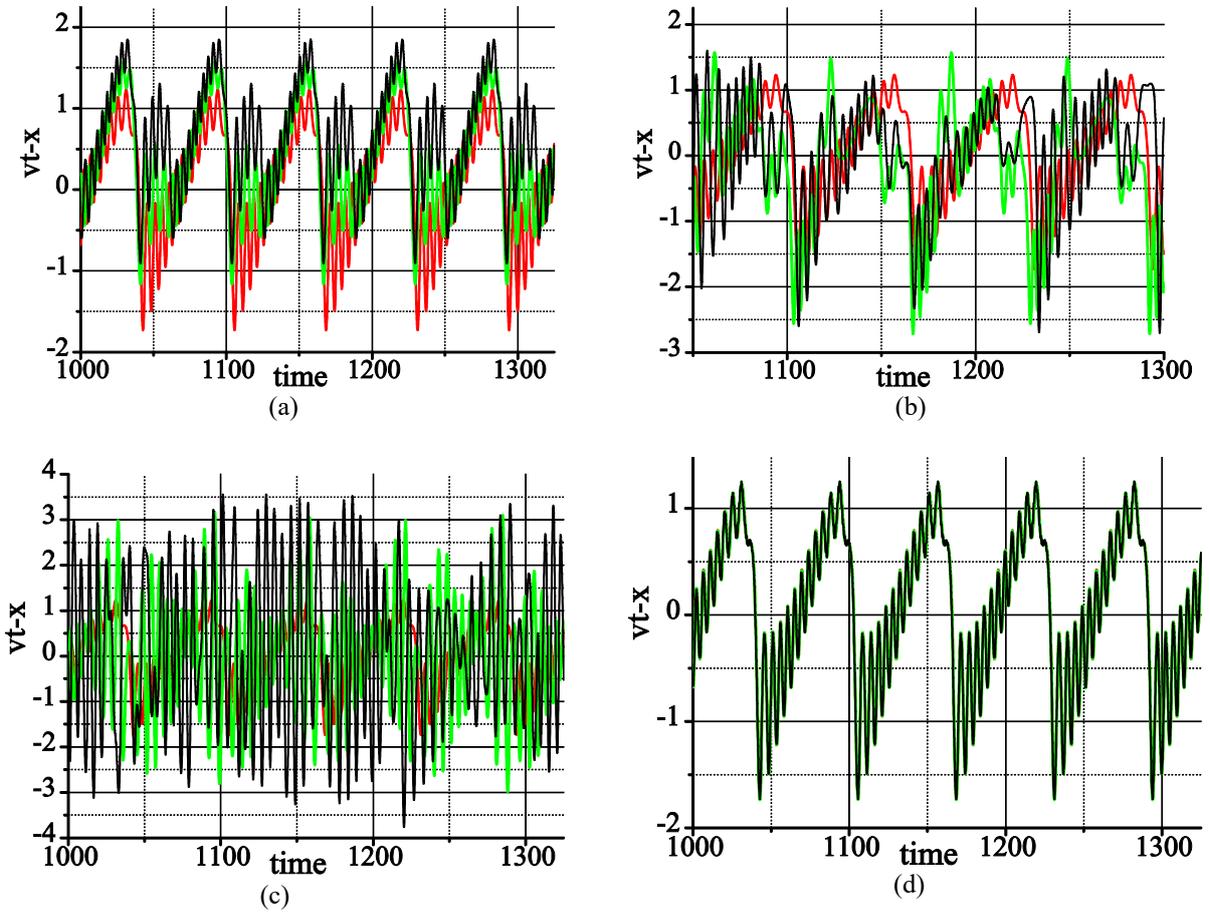

FIG. 5. Time series for the friction force $f_{frict}(t) = vt - x(t)$. Perturbation frequency $\omega = v = 0.1$ (a), $\omega = 2v = 0.2$ (b), $\omega = 1$ (c), $\omega = 8$ (d). Amplitudes are $A = 0$ (red curve), $A = 1$ (green curve), $A = 2$ (black curve).

The same characteristic period $T = 2\pi/v$ is observed for the time series of the friction force for oscillator without perturbation (red curve) and in presence of perturbation with low frequencies, as well as for very high perturbation frequencies. The spectra for the friction force are similar in low and high perturbation frequency limit, i.e. for $\omega \approx v$ and $\omega \gg v$. The frequency spectrum consists mainly of harmonics with frequencies $w = nv$, where $n = 1,2,..$ and major spectral component is at frequency $\omega \approx v$. At $\omega = 1$ for certain amplitudes of perturbation $A$ the oscillator's motion becomes irregular. The frequency composition of oscillation spectra seems almost independent on the perturbation amplitude $A$ for small frequencies of perturbation, but the phases and the amplitudes of constituent harmonic waves are strongly depend on $A$ (except for perturbation with very high frequency ($\omega = 8$) to which the oscillator is not sensitive). In the intermediate case $\omega \approx 1 \div 3$ the spectra includes large number of subharmonics and parametric frequencies with maximum of distribution close to oscillator eigenfrequency $\omega_0 = 1$. Nevertheless even in this intermediate region, provided that the perturbation amplitude is small, the mode-locked regime of oscillations occurs for certain frequencies.

There are two different regimes of temporal behavior of the friction force can be realized by variation of the perturbation amplitude or frequency. The former regime is characterized by irregular time series of the friction force $f_{frict} = vt - x(t)$, and its frequency spectrum contains a large number of spectral components. In this regime the average friction force, which is proportional to the area under the curve $vt - x(t)$, could become very small if frequency components with particular phases compensate constant term in a frequency spectrum over a long time interval.

Another type of oscillator behavior is characterized by oscillations at sum and parametric frequencies of driving and perturbation, so called mode-locked regime where harmonics of the frequency equal to driving parameter $v$ and frequencies $\omega \pm nv$ are dominant. In this case average of the friction force can be substituted by average over period of the lowest frequency component dominating in the spectrum. Oftentimes this frequency is $v$ and the corresponding period is $T = 2\pi/v$. For the mode locked regime the resonant behavior of the friction force could be observed at the frequencies of locking. The friction force can attain large positive or negative values, this corresponds to positive or negative area under the curve $vt - x(t)$ over the period. In the mode-locked regime the positive and negative inputs over the period could be equal, so the total friction force during the period tends to zero. In the time series of the friction force $f_{frict} = vt - x(t)$ for the mode locked regime with high average friction force it appears, as shown in the Fig. 5(a), that in the case of perturbation with $A = 2$, $\omega = v = 0.1$ (black curve) the area of positive part of $f_{frict} = vt - x(t)$ is large comparing to the area of negative part of $f_{frict} = vt - x(t)$. The situation is opposite for the negative values of average friction force, as shown by black curve in the Fig. 5(b) for perturbation with $A = 2$, $\omega = 2v = 0.2$. From the Fig. 5 we could conclude that it is appropriate to approximate $x(t)$ with the set of periodic functions for the case of no perturbation, for the case when perturbations are of small and very high frequencies, as well as for low-amplitude perturbations. Particularly, the time series for $x(t)$ for high frequencies of perturbation are the same as for oscillator without perturbation.

### C. Mode-locked regime of oscillations

#### *1. Analytic approximation of time series and estimation of friction force for unperturbed oscillator*

To illustrate the mode-locked regime let us consider unperturbed oscillator equation Eq. (5) where oscillator eigenfrequency $\omega_0 = 1$ and perturbation amplitude $A = 0$

$$\ddot{x} + \gamma \dot{x} + x - vt + \sin x = 0. \tag{12}$$

Average friction force is defined as following:

$$F_{frict} = \frac{1}{T}\int_0^T (vt - x(t))dt = \frac{1}{T}\int_0^T (\ddot{x} + \gamma\dot{x} + \sin x)dt = \frac{1}{T}\left(\dot{x}|_0^T + \gamma x|_0^T + \int_0^T \sin x(t)\, dt\right). \tag{13}$$

To calculate average friction force we need to know the time dependence (time series) of the coordinate $x(t)$. As follows from the Fig. 5 (red curve serves for $A = 0$), the function $vt - x(t)$ is periodic and the largest period is $T = 2\pi/v$. To find appropriate approximation with periodic func-tions for numerically obtained time series, frequency spectrum of the function $vt - x(t)$ is analyzed. In the Fig. 6(a) the spectrum of the function $f_{frict} = vt - x(t)$ is presented together with corresponding approximation for numerically obtained time series shown in the Fig. 6(b). The presence of nonlinear term $\sin x$ leads to appearance of frequency components $\omega = nv, n = 0, 1, 2, ..$ up to $n = 15$. These harmonics are multiples of frequency equal to driving parameter $v$. Nevertheless the dominant frequency component is $v$. The low-order harmonics of the frequency equal to $v$ provide significant input into average friction force if there is no external perturbation, or perturbation occurs at high frequency far from oscillator's eigenfrequency.

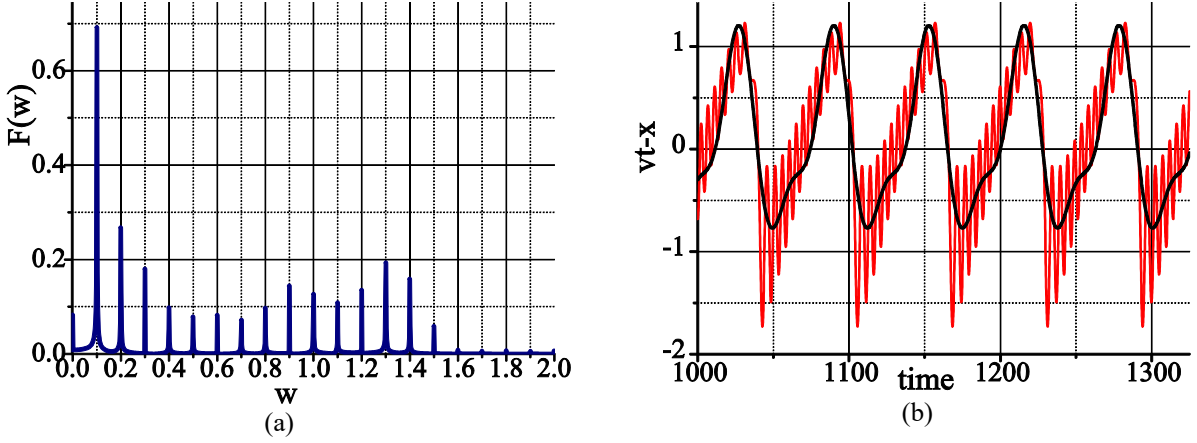

FIG. 6. (a) Frequency spectrum of the friction force $f_{frict} = vt - x(t)$ when there is no perturbation applied ($A = 0$). (b) Time series and its analytical approximation for the friction force. Black curve corresponds to approximation with $x(t) = -0.082 + vt - 0.85\sin(vt + \varphi_1) + 0.32\sin 2vt$, the phase is $\varphi_1 = -0.36$. Red curve serves for the numerically obtained time series $vt - x(t)$.

The approximation for the function

$$f_{frict} = vt - x(t) = 0.082 + 0.85 \sin(vt + \varphi_1) - 0.32 \sin 2vt$$

shown with solid black curve in the Fig. 6(b) is based on frequency spectrum illustrated in the Fig. 6(a) and accounts for the first and second harmonics of driving frequency $v$ only. Approximation

$$x(t) = x_0 + vt + \alpha_1 \sin(vt + \varphi_1) + \alpha_2 \sin 2vt = 0.082 - 0.85 \sin(vt + \varphi_1) + 0.32 \sin 2vt,$$

where coefficients $\alpha_1$ and $\alpha_2$ are chosen equal to the amplitudes of corresponding spectral components, fits visually to time series. Nevertheless the phases of harmonics are also important to obtain correct analytical estimation of friction force. Accounting for it we use the following approximation for the oscillator's coordinate:

$$x(t) = x_0 + vt + \alpha_1 \sin(vt + \varphi_1) + \alpha_2 \sin(2vt + \varphi_2). \tag{14}$$

This approximation is applicable in the mode-locked regime, provided that the spectrum of the friction force consists mainly of first, second harmonics of $v$ and the constant component (as in the Fig. 6). Though this approximation does not include other harmonics, it is appropriate if to choose properly the phases of spectral components. From approximation of Eq. (14) the average friction force can be calculated as following:

$$F_{frict}(A = 0) = \frac{1}{T}\int_0^T (\ddot{x} + \gamma\dot{x} + \sin x)dt =$$

$$\gamma v + \frac{1}{T}\int_0^T \sin(x_0 + vt + \alpha_1\sin(vt + \varphi_1) + \alpha_2\sin(2vt + \varphi_2))dt, \qquad (15)$$

and the similar relation is:

$$F_{frict}(A = 0) = \frac{1}{T}\int_0^T (vt - x)dt = -x_0.$$

In the approximation of Eq. (14) the input into average friction force over the longest period is nonzero for nonlinear crystal lattice potential. Using the first order approximations [13]

$$\sin(I\sin\theta) \approx 2J_1(I)\sin\theta, \quad \cos(I\sin\theta) \approx J_0(I),$$

the average friction force for approximation of Eq. (14) is estimated by the following formula (see Appendix B for the details):

$$F_{frict}(A = 0) = \gamma v - J_1(\alpha_1)\big(J_0(\alpha_2)\sin(x_0 - \varphi_1) + J_1(\alpha_2)\sin(x_0 + \varphi_1 - \varphi_2)\big). \qquad (16)$$

For the mode amplitudes small enough $J_0(x) \approx x/2, J_1(x) \approx 1 - x^2/2$, then

$$F_{frict}(A = 0) \approx \gamma v - (1 - \alpha_1^2/2)\big(0.5\alpha_2\sin(x_0 - \varphi_1) + (1 - \alpha_2^2/2)\sin(x_0 + \varphi_1 - \varphi_2)\big) \approx$$
$$\approx \gamma v - 0.5\alpha_2\sin(x_0 - \varphi_1) - (1 - 0.5(\alpha_1^2 + \alpha_2^2))\big(\sin(x_0 + \varphi_1 - \varphi_2)\big).$$

Equating terms at zero frequency,

$$x_0 + \gamma v - J_1(\alpha_1)\big(J_0(\alpha_2)\sin(x_0 - \varphi_1) + J_1(\alpha_2)\sin(x_0 + \varphi_1 - \varphi_2)\big) = 0, \qquad (17)$$

we could obtain the phases $\varphi_{1,2}$ to satisfy this equation. For example at the Fig. 6(b) the relation $x(t) = -0.082 + vt - 0.85\sin(vt + \varphi_1) + 0.32\sin(2vt + \varphi_2)$ is used to approximate the time series obtained in numerical simulation of the Eq. (5). The amplitudes of the first two harmonics are taken from the spectrum. The positive part of the area under the curve of friction force time series is larger than the negative part. For example in the approximation used to illustrate the time series in the Fig. 6 (b) (the amplitudes of the harmonics are taken from the spectrum and $\varphi_1 = -0.36$ with $\varphi_2 = 0.21$ are set to comply with Eq. (17). For these phases the value of average friction force calculated from Eq. (12) coincides with numerically obtained value. The introduced phases $\varphi_1$ and $\varphi_2$ do not change significantly the fitting function but they are needed to match Eq. (17) and to obtain correct value of the friction force.

It was shown in the numerical experiment that the spectral pikes are observed in the frequency spectrum if perturbation frequency close to driving parameter $\omega \approx v$, or exceeds oscillator eigenfrequency, i.e. $\omega \gg \omega_0$, particularly at $\omega > 5$ ($v = 0.1$ is set in the simulations, $\omega_0 = 1$). These pikes correspond to the frequencies multiple to $v$, it means that harmonics of driving frequency are generated. There are no sum or parametric frequencies $\omega \pm nv$ observed in the spectrum for perturbation frequencies $\omega > 5$ where $\omega \gg v$, as perturbation frequency far from resonance with driving frequency. It was shown numerically in Figs. (1), (3), (4) that average friction force for very high frequencies ($\omega > 5$) is the same as for oscillator without perturbation, and it does not depend on the perturbation amplitude up to $A = 2$.

In presence of external perturbation the average friction force at $\omega_0 = k_1 = x_1 = 1$ is calculated as following:

$$F_{frict} = \frac{1}{T}\int_0^T(vt - x)dt = \frac{1}{T}\int_0^T(\ddot{x} + \gamma\dot{x} + (1 + A\sin\omega t)\sin x)dt. \qquad (18)$$

In averaging of the term $A\sin\omega t \sin x$ it appears that in the approximation of $x(t)$ given by Eq. (14), the additional input into friction force due to the presence of external perturbation $A\sin\omega t$ arises only for certain frequencies, which are $\omega = v, 2v, 3v$ and higher order multiples of $v$. At frequencies of subharmonics $\omega = v/3, v/2$ the average friction force is also affected by the presence of perturbation, but these frequency components are not included into approximation. Other frequency components do not provide input into average friction force.

Particularly, average friction force in the approximation of the Eq. (14) for perturbation with frequency $\omega = v$ is:

$$F_{frict}(\omega = v) = \gamma v - J_1(\alpha_1)\big(J_0(\alpha_2)\sin(x_0 - \varphi_1) + J_1(\alpha_2)\sin(x_0 + \varphi_1 - \varphi_2)\big) + \\ +0.5AJ_0(\alpha_1)\big(J_0(\alpha_2)\cos x_0 + J_1(\alpha_2)\cos(x_0 - \varphi_2)\big). \quad (19)$$

$$F_{frict}(\omega = v) \approx \gamma v - (1 - \alpha_1^2/2)\big(0.5\alpha_2\sin(x_0 - \varphi_1) + (1 - \alpha_2^2/2)\sin(x_0 + \varphi_1 - \varphi_2)\big) + \\ +0.25A\alpha_1\big(0.5\alpha_2\cos x_0 + (1 - \alpha_2^2/2)\cos(x_0 - \varphi_2)\big) \approx \\ \approx \gamma v - 0.5\alpha_2\sin(x_0 - \varphi_1) - \big(1 - 0.5(\alpha_1^2 + \alpha_2^2)\big)\sin(x_0 + \varphi_1 - \varphi_2) + \\ +0.125A\alpha_1\alpha_2\cos x_0 + 0.25A\alpha_1\cos(x_0 - \varphi_2).$$

## *2. Input into friction force for perturbation at frequency of driving force*

Mode-locked regime of oscillations was observed in numerical experiment when perturbation occurs at the frequency $\omega = v$ of driving term in oscillator equation. The value of the average friction force is estimated by the Eq. (19) which is more precise for perturbations of small amplitude when the inputs of generated first and second harmonics of $v$ into friction force are most important, as shown in the spectrum in the Fig. 7(a). Constant term at zero spectral frequency is increased twice comparing with unperturbed oscillator. Computation of the friction force spectrum shows that the amplitude of the second harmonic of driving frequency in frequency spectrum increases with growth of perturbation amplitude $A$ and the first harmonic's amplitude drops correspondingly. Increase of perturbation amplitude $A$ leads to more efficient generation of the second as well as higher-order harmonics at the expense of the first harmonic. Thus instead of approximation with Eq. (14) it should be used another approximation with high-order harmonics of $v$ which would be more appropriate for large perturbation amplitudes.

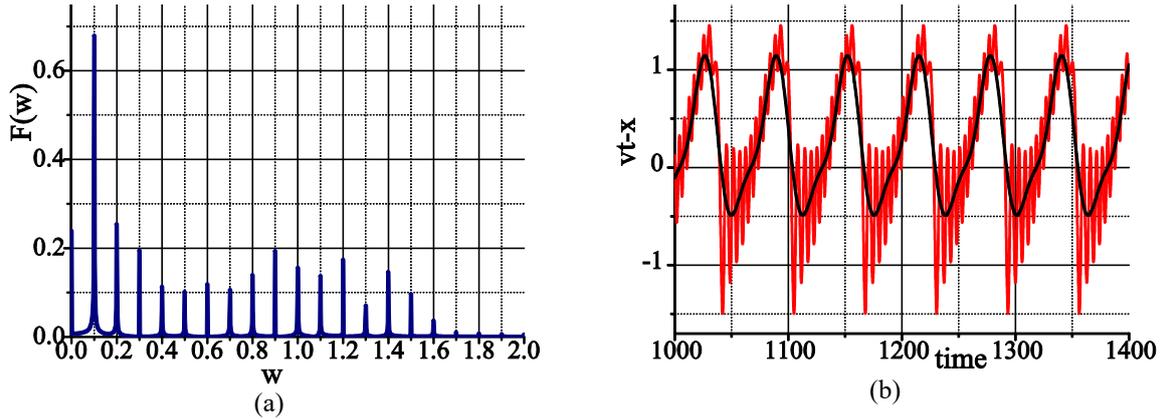

FIG. 7. a) Frequency spectrum of function $f_{frict} = vt - x(t)$ for $\omega = v$, $A = 0.5$. b) Time series and its analytical approximation for friction force. Black curve corresponds to the approximation with $x(t) = -0.239 + vt - 0.73\sin(vt + \varphi_1) + 0.22\sin 2vt$ and $\varphi_1 = -0.3$. Red curve serves for the numerically obtained time series.

To calculate the value of the friction force from Eq. (19) the perturbation amplitude is taken as $A = 0.5$. In the approximation used for the Fig. 7(b) $x(t) = -0.239 + vt - 0.73\sin(vt + \varphi_1) + 0.22\sin 2vt$ with the amplitudes of first and second harmonics of driving parameter $v$ corresponding to the spectrum in the Fig. 7(a). From Eq. (19) it is follows that $F_{frict}(\omega = v, A = 0.5) \approx$ 0.239, this value coincides with numerically calculated average friction force for the phases $\varphi_1 = -0.3$ and $\varphi_2 = 0$ obtained by equating terms with zero frequency.

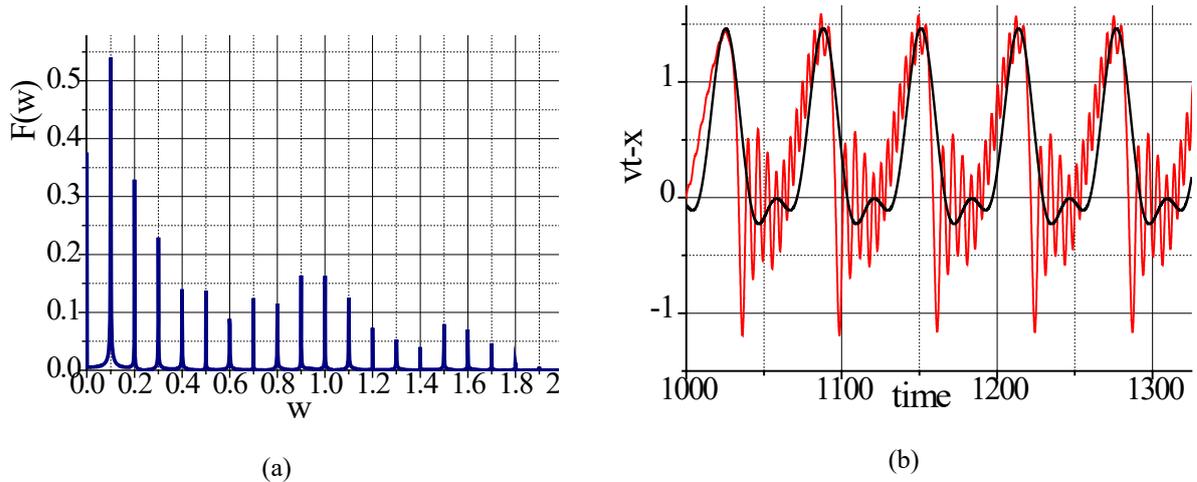

(a)             (b)

FIG. 8. a) Spectrum of the friction force $vt - x(t)$ for $\omega = v$, $A = 1$. b) Time series and its analytical approximation. Black curve corresponds to the approximation with $x(t) = -0.375 + vt - 0.74\sin(vt + \varphi_1) - 0.35\sin(2vt + \varphi_2)$ and $\varphi_1 = -0.37$, $\varphi_2 = -2.5$. Red curve serves for the numerically obtained time series.

Increase for perturbation amplitude up to $A = 1$ results in generation of high order harmonics of driving parameter $v$, as shown in the Fig. 8(a). Spectral amplitude of the third harmonic becomes comparable with second harmonic, higher-order components are also increased comparing to Figs. 6(a)-7(a).

In the approximation used for perturbation with $A = 1$, $\omega = v$ we again account only 2 harmonics, as shown in Fig. 8 (b), thought this approximation is rough for the frequency spectrum shown in Fig. 8(a). In the Fig. 8(b) for the time series of friction force (red curve) the approximation

$$f_{frict} = vt - x(t) = 0.375 + 0.74\sin(vt - 0.37) + 0.35\sin(2vt - 2.5)$$

is shown by black curve using values of the amplitudes of first and second harmonics from the frequency spectrum and selecting phases as $\varphi_1 = -0.37$, $\varphi_2 = -2.5$ to match the amplitude equations. From Eq. (19) it is follows that $F_{frict}(\omega = v, A = 1) \approx 0.375$, this value close to numerically obtained value of friction force defined by the amplitude of zero frequency spectral component in Fig. 8(a).

In averaging of the term $A\sin\omega t \sin x$ it appears that in the chosen approximation of $x(t)$ given by Eq. (9), the additional input into the friction force due to the presence of external perturbation $A\sin\omega t$ arises only for certain frequencies, which are $\omega = v, 2v, 3v$ and higher order multiples of $v$. At subharmonics $\omega = v/3, v/2$ the average friction force is also affected by the presence of perturbation. For other frequencies averaging gives zero value of input into friction force. Approximation by Eq. (14) shows that the average friction force is strongly affected by the presence of external perturbation at the condition that the frequency of this perturbation is a multiple of driving parameter $v$. The similar behavior is illustrated in the Fig. 1 for the friction force dependence on perturbation frequency. Approximation for $x(t)$ with Eq. (14) is suitable for low perturbation frequencies when the spectra of $x(t)$ can be approximated by first and second harmonics of frequency $v$ together with the constant term [Fig. 7(a)].

### *3. Input into friction force for perturbation at frequency of second harmonic of driving force*

When perturbation occurs at frequency $\omega = 2v$, the second harmonic of driving parameter $v$, spectrum of the friction force [Fig. 9(a)] contains first and third harmonics of $v$, also subharmonics $v(n + 1)/2$ of $v$ are generated. Increase for the amplitude of perturbation with frequency $\omega = 2v$ leads to amplification of spectral components $\omega = v, 3v$ of friction force [Fig. 9(a)]. The amplitude of spectral component with frequency $v$ is large comparing to spectral component at frequency $3v$. At perturbation with amplitude $A = 0.5$ the ratio of these components's amplitudes is $\alpha_{\omega-v=v}/\alpha_{\omega+v=3v} = (\omega + v)/(\omega - v) = 3$. This value coincides with analytical formula of Eq. (31) obtained in Appendix C in Eq. (C14) for estimation of amplitudes of anti-Stokes and Stokes frequency components $\omega \pm v$ resulting from parametric interaction of driving mode with frequency $v$ with small amplitude perturbation with frequency $\omega$.

Another approximation that could be used instead of Eq. (14) is

$$x(t) = x_0 + vt + \alpha_1 \sin(vt + \varphi_1) + \alpha_3 \sin 3vt \tag{20}$$

and average friction force is calculated as following:

$$F_{frict}(\omega = 2v) = \\ = \gamma v - J_1(\alpha_1)J_0(\alpha_3)\sin(x_0 - \varphi_1) + \frac{A}{2}(J_1(\alpha_1)J_0(\alpha_3)\cos(x_0 + \varphi_1) + J_0(\alpha_1)J_1(\alpha_3)\cos x_0). \tag{21}$$

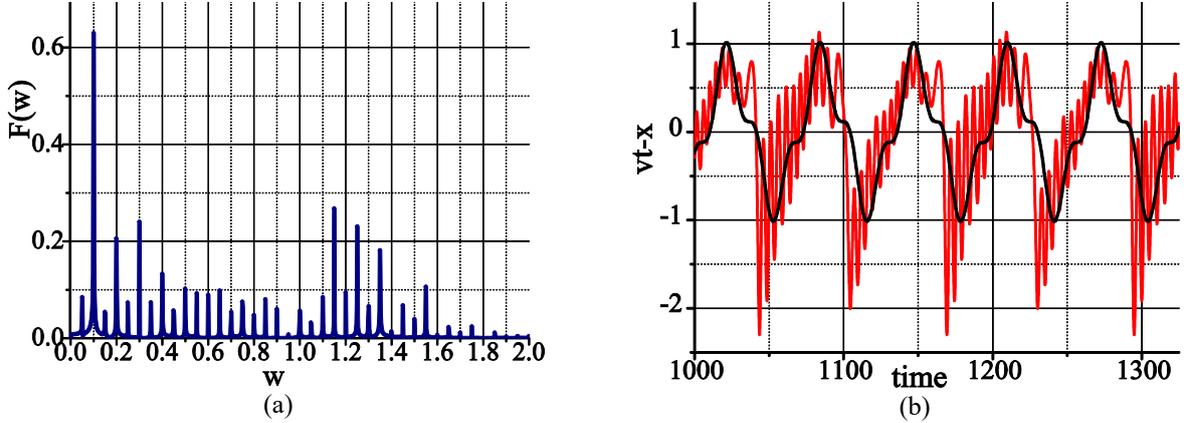

FIG. 9. Frequency spectrum (a) of function $vt - x(t)$ for perturbation with $A = 0.5$ and $\omega = 2v$. (b) Time series and its analytical approximation. Black curve corresponds to time series approximation of $vt - x(t)$ by $x(t) = 0.001 + vt - 0.77\sin(vt + \varphi_1) + 0.25\sin 3vt$ with phase $\varphi_1 = -0.154$. Red curve serves for the numerically obtained time series $vt - x(t)$.

To calculate the value of the friction force from Eq. (21) the perturbation amplitude is taken as $A = 0.5$. In the approximation $x(t) = 0.001 + vt - 0.77\sin(vt + \varphi_1) - 0.25\sin 3vt$ used for the Fig. 9(b) with the amplitudes of the first and the third harmonics corresponding to spectral components shown in the Fig. 9(a). From Eq. (21) it is follows that $F_{frict}(\omega = 2v, A = 0.5) \approx -0.001$, this value coincides with numerically calculated average friction force for the phase $\varphi_1 = -0.154$ obtained by equating terms with zero frequency. The areas of positive and negative parts of time series of function $f_{frict} = vt - x(t)$ over period in this approximation are close; it corresponds to very small friction force.

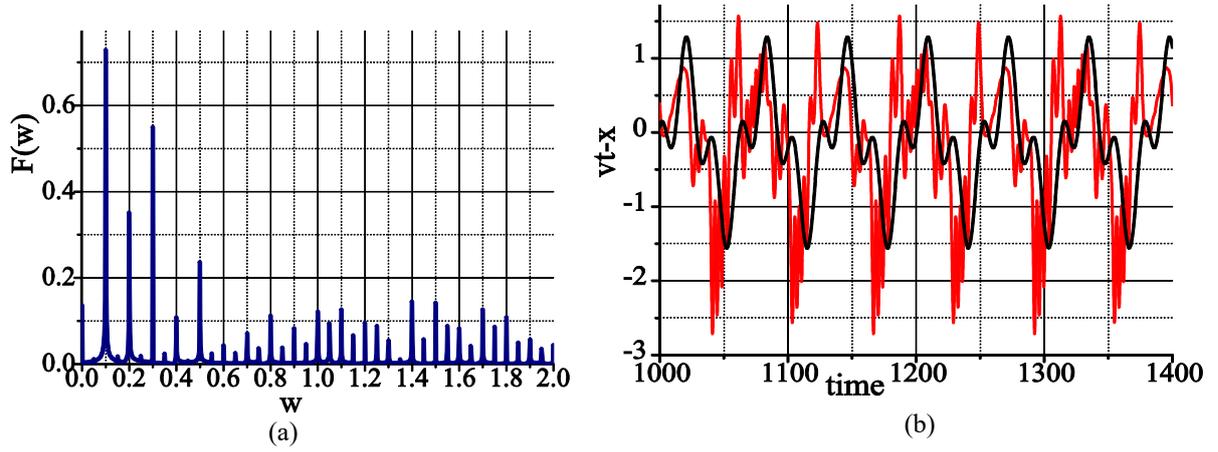

FIG. 10. (a) Frequency spectrum (a) of function $vt - x(t)$ for perturbation with $A = 2$ and $\omega = 2v$. Time series and its analytical approximation are shown in the panel (b). Black curve corresponds to time series approximation of $vt - x(t)$ by $x(t) = 0.135 + vt - 0.88\sin(vt + \varphi_1) + 0.55\sin 3vt$ with phase $\varphi_1 = 0.13$. Red curve serves for the numerically obtained time series $vt - x(t)$.

Value of the friction force for the perturbation amplitude $A = 2$ is calculated in the same way as for $A = 0.5$ using time series approximation $x(t) = 0.135 + vt - 0.88\sin(vt + \varphi_1) + 0.55\sin 3vt$ shown in the Fig. 10(b) with the amplitudes of the first and the third harmonics of $v$ corresponding to the spectrum in the Fig. 10(a). Calculated from Eq. (21) average friction force $F_{frict}(\omega = 2v, A = 2) \approx -0.135$, this value agrees with numerically calculated average friction force for the phase $\varphi_1 = 0.13$ obtained from equating terms with zero frequency. The area of approximating function over period is negative and average friction force is negative.

More accurate, but much more sophisticated expression for the friction force can be obtained if to account for the third harmonic and the phases of all harmonics, i.e. instead of approximations given by Eq. (14) or Eq. (20) use the following:

$$x = x_0 + vt + \alpha_1\sin(vt + \varphi_1) + \alpha_2\sin(2vt + \varphi_2) + \alpha_3\sin(3vt + \varphi_3). \quad (22)$$

In the Appendix B the approximation of Eq. (22) is considered for all phases set to zero.

### *4. Analytic estimation of friction force in three-mode approximation with perturbation frequency mode and its satellite anti-Stokes and Stokes modes*

In numerical simulation resonant behavior of average friction force has been observed at perturbation frequencies higher than driving frequency. In numerically obtained amplitude spectra the amplitudes of low-order harmonics of $v$ decay with increase of perturbation frequency $\omega$, where $v < \omega < 5$, but frequencies $\omega \pm nv$ are generated. On the basis of spectral analysis of time series of the friction force it could be used a different approximation for $x(t)$ at higher perturbation frequencies. The presence of the perturbation at the frequency $\omega$ in the mode locked regime leads to generation of the side harmonics $\omega \pm v, \omega \pm 2v, ...$ To get qualitative description of average friction force for this case we consider the simplified approximation and calculate the input from central spectral component at frequency $\omega$ of perturbation and satellite anti-Stokes and Stokes frequencies $\omega + v$, $\omega - v$.

Using approximation
$$x = x_0 + vt + \alpha_{\omega-v}\sin(\omega - v)t + \alpha_\omega \sin \omega t + \alpha_{\omega+v}\sin(\omega + v)t, \quad (23)$$

in averaging of Eq. (18) we suppose that $\omega$ is a multiple of $v$, i.e. $\omega = nv$, and $\omega > v, 2v, 3v$. Expression for average friction force at $k_1 = x_1 = 1$ (see Appendix C) is following:

$$F_{frict}(\omega) = \frac{1}{T}\int_0^T (\ddot{x} + \gamma\dot{x} + (1 + A\sin\omega t)\sin x)dt =$$
$$= \gamma v + (-\sin x_0 J_1(\alpha_\omega) + 1/2 A J_0(\alpha_\omega)\cos x_0\,)(J_0(\alpha_{\omega-v})J_1(\alpha_{\omega+v}) + J_0(\alpha_{\omega+v})J_1(\alpha_{\omega-v}))$$
(24)

$$\sin x_0 = -\frac{\alpha_1 \gamma v}{(J_0(\alpha_1)J_0(\alpha_3) - J_1(\alpha_1)J_1(\alpha_3))(J_0(\alpha_2) - J_1(\alpha_2))}.$$
(25)

Average friction force for very high perturbation frequencies ($\omega > 5$ for the Fig. 1) does not depend on perturbation amplitude, as for the components $\omega$, $\omega \pm v$ their amplitudes $\alpha_\omega$, $\alpha_{\omega\pm v}$ are negligible (as it was found from the amplitude spectrum). But at $\omega \approx v$ (as well as at $\omega = 0$) the components $v, 2v, 3v$ are considerable and they give an extra input into the friction force, which is not taken into account in Eq. (24). This extra input, which does not depend on the amplitude of perturbation, could be calculated in the approximation of Eq. (14). Approximate value of the friction force is the same as in numerical simulation of oscillator without perturbation.

In the absence of side harmonics in approximation $x = x_0 + vt + \alpha_\omega \sin\omega t$ the signal at frequency $\omega$ is not giving an input into friction force. In the simplified approximation without central frequency component $\omega$

$$x = x_0 + vt + \alpha_{\omega-v}\sin(\omega - v)t + \alpha_{\omega+v}\sin(\omega + v)t. \quad (26)$$

Substitution of Eq. (26) into Eq. (5) and averaging using Eq. (6) results in following:

$$F_{frict}(\omega) = \frac{1}{T}\int_0^T (\ddot{x} + \gamma\dot{x} + (1 + A\sin\omega t)\sin x)dt \approx$$
$$\approx \gamma v + \frac{A}{2}\cos x_0\,(J_0(\alpha_{\omega-v})J_1(\alpha_{\omega+v}) + J_0(\alpha_{\omega+v})J_1(\alpha_{\omega-v}))$$
(27)

In this derivation we suppose that $\omega$ is a multiple of $v$. To calculate $\cos x_0$ and the coefficients $\alpha_{\omega\pm v}$ let us collect the terms at the frequencies $\omega \pm v$ in the following equation:

$$\left(\alpha_{\omega\pm v}(1 - (\omega \pm v)^2) + \frac{A}{2}\sin x_0\,(J_0(\alpha_{\omega+v})J_0(\alpha_{\omega-v}) - 2J_1(\alpha_{\omega+v})J_1(\alpha_{\omega-v}))\right)\sin(\omega \pm v)t +$$
$$+ \left(\gamma(\omega \pm v)\alpha_{\omega\pm v} \mp \frac{A}{2}J_0(\alpha_{\omega+v})J_0(\alpha_{\omega-v})\cos x_0\right)\cos(\omega \pm v)t = 0$$
(28)

The expressions to define $\cos x_0$ and the amplitudes of the side spectral components of $\omega$ are:

$$\cos x_0 = \pm\frac{2\gamma(\omega\pm v)\alpha_{\omega\pm v}}{AJ_0(\alpha_{\omega+v})J_0(\alpha_{\omega-v})}, \quad (29)$$

$$\alpha_{\omega\pm v} = -\frac{A\sin x_0(J_0(\alpha_{\omega+v})J_0(\alpha_{\omega-v}) - 2J_1(\alpha_{\omega+v})J_1(\alpha_{\omega-v}))}{2(1 - (\omega\pm v)^2)}, \quad (30)$$

where

$$\frac{\alpha_{\omega+v}}{\alpha_{\omega-v}} = -\frac{\omega - v}{\omega + v}. \quad (31)$$

Signs of spectral components $\omega \pm v$ are opposite. And for $\omega > v$ $|\alpha_{\omega+v}| < |\alpha_{\omega-v}|$.

For example the ratio between the amplitudes of side harmonics is calculated for Figs. 8-9 and Fig. 11 for perturbation frequencies $\omega = 2v, \omega_0$ is in a good agreement with formula of Eq. (31). Substitution of the Eq. (29) into Eq. (27) results in formula for estimation of average friction force in approximation of Eq. (26):

$$F_{frict} = \gamma\left(v \pm \frac{(\omega\pm v)\alpha_{\omega\pm v}(J_1(\alpha_{\omega+v})J_0(\alpha_{\omega-v}) + J_1(\alpha_{\omega-v})J_0(\alpha_{\omega+v}))}{J_0(\alpha_{\omega+v})J_0(\alpha_{\omega-v})}\right). \quad (32)$$

From this expression it is follows that the friction force in the oscillator model could be negative.

### D. Irregular regime of oscillations

At the Fig. 11 the perturbation frequency coincides with oscillator's eigenfrequency $\omega = \omega_0 = 1$. Amplitude of spectral component of friction force with frequency $v$ as well as of its harmonics is small and drops with the increase of the perturbation amplitude, so interaction of perturbation with the external driving force becomes weak. The spectral distribution is not symmetric relatively $\omega_0 = 1$ and depends on the amplitude of perturbation. For $A = 0.5$ the dominant frequency components are $\omega \pm v$, $\omega \pm 2v$, for $A = 1$ the dominant frequency components are $\omega \pm v$ ($\omega = 0.9$ and $\omega = 1.1$), and the amplitude of the lower spectral component $\omega = 0.9$ is higher than that of spectral component $\omega = 1.1$. From the same formula of Eq. (31) the ratio between the amplitudes of these components is estimated:

$$\alpha_{\omega-v=9v=0.9}/\alpha_{\omega+v=11v=1.1} = (\omega+v)/(\omega-v) \approx 1.2. \tag{33}$$

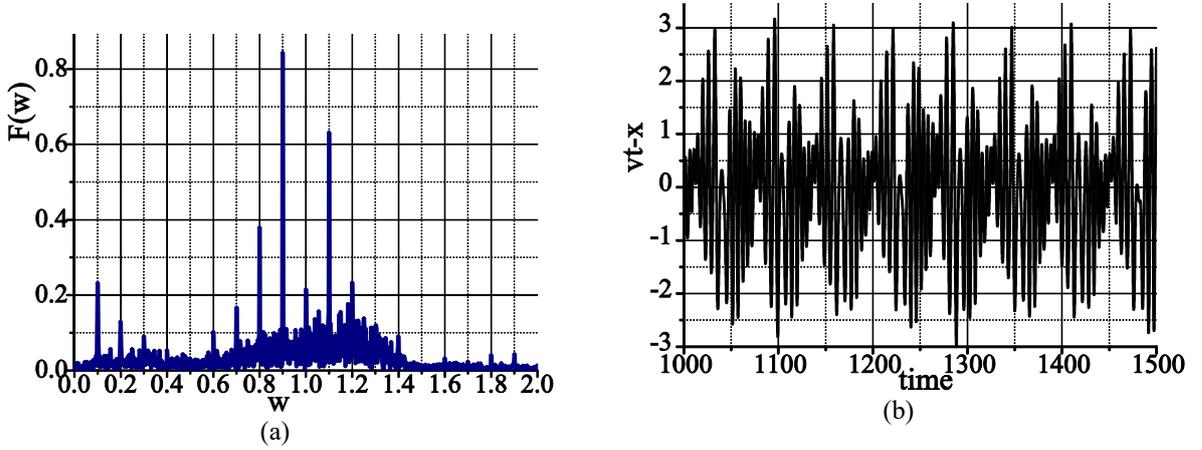

FIG. 11. Frequency spectrum (a) and time series (b) for the friction force in the presence of perturbation with frequency $\omega = 1$ and amplitude $A = 1$.

From numerically obtained spectrum in the Fig. 11(a) it could be obtained that ratio $\alpha_{\omega-v=0.9}/\alpha_{\omega+v=1.1} \approx 1.2$. This value well coincides with value of ratio $\alpha_{\omega-v}/\alpha_{\omega+v}$ calculated in Eq. (33) using approximation of Eq. (31) because spectral component with central frequency $\omega = 1$ is small. At perturbation frequency $\omega = \omega_0 = 1$ there is no characteristic period can be determined from the time series and wide amplitude spectrum. Thus it is difficult to propose an analytical approximation of $x(t)$ in this case. Nevertheless, in the case of large spectrum with many frequencies it could happens that spectral components compensate each other over averaging interval. Thus it would results in zero value of average friction force. For example average friction force for the perturbation with $\omega = 1$ and $A = 1$ is small, spectrum and time series for the friction force in this case are shown in the Figs. 11(a) and 11(b) respectively.

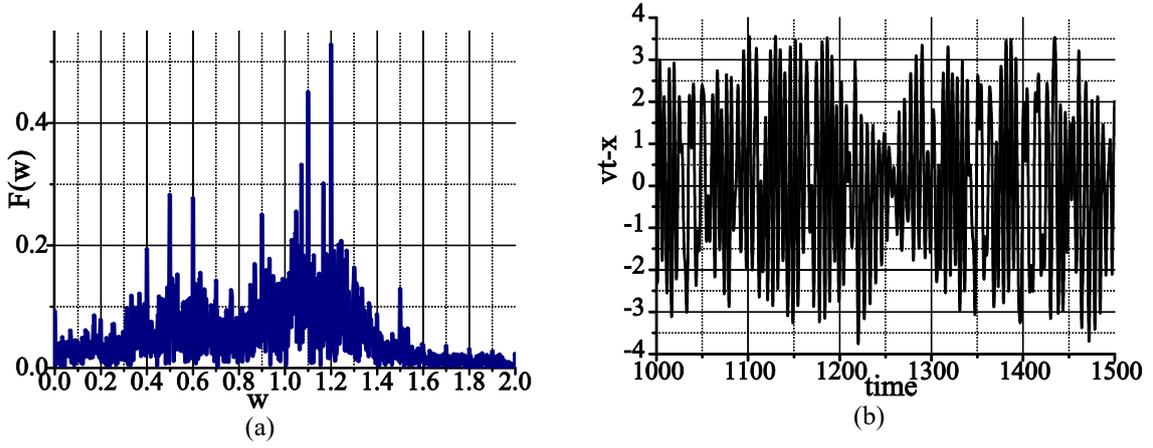

FIG. 12. Frequency spectrum (a) and time series (b) for the friction force in the presence of perturbation with frequency $\omega = 1$ and amplitude $A = 2$.

If perturbation with the same frequency $\omega = 1$ and increased amplitude $A = 2$ is applied to oscillator, the frequency spectrum at Fig. 12(a) is also wide and contains a lot of frequency components. However they do not compensate each other and the average friction force is noticeable. Irregular time series of the function $vt - x(t)$ are illustrated at Fig. 12(b).

## IV. DISCUSSION

In the mode-locked regime the friction force is approximated with a sum of periodic components with different frequencies using frequency spectrum of oscillating process's time series obtained in numerical simulation. For the case of no perturbation applied to oscillator it was used one-mode approximation of Eq. (9), however it provides less value of average friction force comparing to numerical integration. Using two-mode approximation of Eq. (14) to estimate average friction force in Eq. (16) it was obtained numerical value equal to average friction force in numerical simulation.

In particular cases of perturbation frequencies $\omega = v, 2v$ there were applied one-mode approximation of Eq. (8) which was used to obtain dependence of friction force in Eqs.(9)-(10) on perturbation amplitude, and two-mode approximations described by Eq. (14) and Eq. (20), where friction force is defined by Eq. (19) and Eq. (21) respectively.

There is a similarity between numerically obtained and analytical approximations for time series and average values for friction forces when perturbation occurs at low or very high frequency comparing to oscillator eigenfrequency. For intermediate frequencies analytical results are close to numerically obtained data when frequency spectrum and the time series for $x(t)$ can be represented by several spectral components.

## V. CONCLUSION

In the numerical experiment we observed two different regimes for friction force time dependence for various amplitudes and frequencies of external perturbation. We discussed these regimes and illustrated the difference between them by representing the time series and frequency spectra for friction force. The first one is the regime of mode locking, it occurs without perturbation due to nonlinearity of potential. Mode-locked regime could also be realized for low-amplitude perturbations and two frequency ranges of low and high frequencies away from oscillator eigenfrequency $\omega_0 = 1$. In this regime friction force spectrum is discrete and consists of limited number of harmonics of $v$ and frequencies $\omega \pm nv$. Average friction force for this regime is estimated by averaging its approximate function which consists of several spectral components. The nonzero input into the friction force, which can be positive or negative, results from averaging of these spectral components appearing due to nonlinearity of periodic lattice potential. In the numerical simulations it was observed the resonant behavior of the friction force at perturbation frequencies proportional to driving parameter $v = 0.1$.

Two different oscillation regimes are observed, they determined by the frequency and amplitude of perturbation. The first one is the regime of mode locking at the frequencies multiple to driving parameter $v$. This occurs either when perturbation frequency $\omega$ is close to frequency equal numerically to driving parameter $v$, either it is considerably large comparing to eigenfrequency $\omega_0$ of the oscillator. Particularly, average friction force attains for its global maximum when the perturbation frequency is exactly equal to $v$. Negative values of average friction force are observed in numerical simulations for certain intervals of perturbation amplitude when the perturbation frequency $\omega = 2v$.

The regime, which can be considered as a partially mode locked, occurs for small amplitude perturbations with frequency located in the intermediate region near the oscillator eigenfrequency $\omega \approx \omega_0$. In this mode locked regime the system follows a force which contains a constant spectral component together with harmonics $v, 2v, 3v$ and so forth of driving parameter $v$. In order to estimate the average friction force in the mode-locked regime analytically, we propose to approximate its temporal profile with a set of harmonics of $v$. This allows to substitute averaging over the full observation interval by averaging over the longest period (which is $T = 2\pi/v$ if no subharmonic of $v$ are generated).

Another propagation regime is characterized by generation of a large number of spectral components far away from driving parameter $v$ and its harmonics. In this regime oscillator does not follow the external driving force, but rather oscillates with frequencies $\omega \pm nv$ and $\omega_0 \pm nv$, which result from interaction of driving mode $v$ with perturbation mode $\omega$ or oscillator eigenmode $\omega_0$. In this case there is no definite period can be determined for the time series of friction force, as oscillator spectrum contains many spectral components. For certain amplitudes and frequencies of perturbation average friction force tends to zero value or becomes negative at time interval of intergation. In irregular regime of oscillations average friction force in numerical experiment tends to zero for certain amplitudes and frequencies of perturbation. This phenomenon could be explained by compensation of all frequency components of the friction force over integration interval.

## ACKNOLEDGEMENT

Elena Kazantseva is appreciating Computer Science and Mathematics Division, Center for Engineering Science Advanced Research (CESAR), Oak Ridge Associated Universities and Oak Ridge Institute for Science and Education (ORAU/ORISE) for hospitality and support.

## APPENDIX A: SINGLE MODE APPROXIMATION OF FRICTION FORCE

Assume that oscillator displacement is approximated as

$$x(t) = vt - \alpha_1 \sin(vt). \tag{A1}$$

In this approximation oscillator evolves at one mode with frequency of driving term and amplitude $\alpha_1$. Averaging over characteristic period $T = 2\pi/v$ leads to following equation for average friction force:

$$F_{frict} = \frac{1}{T}\int_0^{T=2\pi/v}(vt - x(t))dt = \frac{v}{2\pi}\int_0^{2\pi/v}\left(\ddot{x} + \gamma\dot{x} + (1 + A\sin(\omega t))\sin(x)\right)dt =$$
$$= \frac{v}{2\pi}\left[\dot{x}\Big|_0^{2\pi/v} + \gamma x\Big|_0^{2\pi/v} + \int_0^{2\pi/v}\sin(x)\,dt + \frac{A}{2}\int_0^{2\pi/v}(\cos(\omega t - x) - \cos(\omega t + x))dt\right]$$

(A2)

Substitution of $x(t) = vt - \alpha_1\sin(vt)$ and $\dot{x}(t) = v - v\alpha_1\cos(vt)$ into Eq. (A2) results in

$$F_{frict} = \gamma v + \frac{v}{2\pi}\int_0^{2\pi/v}\sin(vt - \alpha_1\sin(vt))\,dt + \frac{vA}{4\pi}\int_0^{2\pi/v}\left(\cos((\omega - v)t + \alpha_1\sin(vt)) - \cos((\omega + v)t - \alpha_1\sin(vt))\right)$$

$$= \gamma v + \frac{vA}{4\pi}\int_0^{2\pi/v}\left(\cos((\omega - v)t + \alpha_1\sin(vt)) - \cos((\omega + v)t - \alpha_1\sin(vt))\right)dt$$

(A3)

If $\omega$ is a multiple of $v$, $\omega = nv$, where $n$ – is integer, Eq. (A3) becomes

$$F_{frict} = \gamma v + \frac{vA}{4\pi}\int_0^{2\pi/v}\left(\cos((n-1)vt + \alpha_1\sin(vt)) - \cos((n+1)vt - \alpha_1\sin(vt))\right)dt.$$

(A4)

From definition of Bessel function [13],

$$2\pi J_n(x) = \int_0^{2\pi}e^{i(x\sin\varphi - n\varphi)}d\varphi = \int_0^{2\pi/v}e^{i(\alpha_1\sin vt - nvt)}dt,$$

(A5)

with $\varphi = vt, x = \alpha_1$,

$$\frac{2\pi}{v}J_n(\alpha_1) = \int_0^{2\pi/v}\cos(nvt - \alpha_1\sin vt)dt, \frac{2\pi}{v}(-1)^n J_n(\alpha_1) = \int_0^{2\pi/v}\cos(nvt + \alpha_1\sin vt)dt.$$

(A6)

From Eq. (A4) it is follows that

$$F_{frict} = \gamma v + \frac{vA}{4\pi}\int_0^{2\pi/v}\left(\cos((n-1)vt + \alpha_1\sin(vt)) - \cos((n+1)vt - \alpha_1\sin(vt))\right)dt =$$
$$\gamma v + \frac{A}{2}\left((-1)^{n-1}J_{n-1}(\alpha_1) - J_{n+1}(\alpha_1)\right).$$

(A7)

Consider examples of oscillator with $\omega_0 = 1$ without perturbation and in presence of perturbation with frequency resonant to driving frequency or its high-order harmonics. In absence of perturbation $n = 0, \omega = 0$,

$$F_{frict} = \gamma v + \frac{vA}{4\pi}\int_0^{2\pi/v}(\cos(vt + \alpha_1\sin(vt)) - \cos(vt + \alpha_1\sin(vt)))dt = \gamma v. \quad (A8)$$

When perturbation occurs at the frequency of driving term $n = 1, \omega = v$,

$$F_{frict}(\omega = v, A) = \gamma v + \frac{A}{2}(J_0(\alpha_1) - J_2(\alpha_1)). \quad (A8)$$

Using recurrence relation

$$J_{n+1}(z) = \frac{2}{z}J_n(z) - J_{n-1}(z), \tag{A9}$$

and setting the amplitude of oscillation mode $\alpha_1 = 1 = z$, we obtain

$$J_{n+1}(1) = 2J_n(1) - J_{n-1}(1), J_2(1) = 2J_1(1) - J_0(1), J_3(1) = 2J_2(1) - J_1(1) = 3J_1(1) - 2J_0(1).$$

Friction force in one mode approximation for perturbation with amplitude $A$ and frequency $\omega = \nu$

$$F_{frict}(\omega = \nu) = \gamma\nu + \frac{\nu A}{4\pi}\int_0^{2\pi/\nu}(\cos(\alpha_1\sin(\nu t)) - \cos(2\nu t - \alpha_1\sin(\nu t)))dt =$$
$$= \gamma\nu + \frac{\nu A}{4\pi}\left(\frac{2\pi}{\nu}J_0(\alpha_1) - \frac{2\pi}{\nu}J_2(\alpha_1)\right) = \gamma\nu + \frac{A}{2}(J_0(\alpha_1) - J_2(\alpha_1)) = \gamma\nu + A(J_0(\alpha_1) - J_1(\alpha_1)). \tag{A10}$$

$$F_{frict}(\omega = \nu, A, \alpha_1 = 1) = \gamma\nu + \frac{A}{2}(J_0(1) - J_2(1)) = \gamma\nu + A(J_0(1) - J_1(1)). \tag{A11}$$

For different perturbation amplitudes $A$ at $\alpha_1 = 1, \gamma = 0.2, \nu = 0.1$,

$$F_{frict}(\omega = \nu, A = 0.5) = \gamma\nu + A(J_0(1) - J_1(1)) \approx 0.02 + 0.16 = 0.18. \tag{A12.1}$$
$$F_{frict}(\omega = \nu, A = 1) = \gamma\nu + A(J_0(1) - J_1(1)) \approx 0.02 + 0.32 = 0.34. \tag{A12.2}$$
$$F_{frict}(\omega = \nu, A = 2) = \gamma\nu + A(J_0(1) - J_1(1)) \approx 0.02 + 0.65 = 0.67. \tag{A12.3}$$

At $\gamma = 0.15$

$$F_{frict}(A = 0.5) = \gamma\nu + A(J_0(1) - J_1(1)) \approx 0.015 + 0.16 = 0.165. \tag{A12.4}$$
$$F_{frict}(A = 1) = \gamma\nu + A(J_0(1) - J_1(1)) \approx 0.015 + 0.32 = 0.335. \tag{A12.5}$$
$$F_{frict}(A = 2) = \gamma\nu + A(J_0(1) - J_1(1)) \approx 0.015 + 0.65 = 0.67. \tag{A12.6}$$

When perturbation is of amplitude $A$ and frequency $\omega = 2\nu$, $n = 2$, average friction force

$$F_{frict}(A = 1, \omega = 2\nu, \alpha_1) = \gamma\nu +$$
$$+ \frac{\nu A}{4\pi}\int_0^{2\pi/\nu}(\cos(\nu t + \alpha_1\sin(\nu t)) - \cos(3\nu t - \alpha_1\sin(\nu t)))dt = \gamma\nu - \frac{A}{2}(J_1(\alpha_1) + J_3(\alpha_1)). \tag{A13}$$

Estimation of friction force in one mode approximation for perturbation with amplitude $A$ and frequency $\omega = 2\nu$ results in

$$F_{frict}(\omega = 2\nu, A, \alpha_1 = 1) = \gamma\nu - \frac{A}{2}(J_1(1) + J_3(1)) = \gamma\nu + A(J_0(1) - 2J_1(1)). \tag{A14}$$

For different perturbation amplitudes $A$ at $\gamma = 0.2$

$$F_{frict}(\omega = 2\nu, A = 0.5) = \gamma\nu + A(J_0(1) - 2J_1(1)) \approx 0.02 - 0.5 * 0.115 = -0.037. \tag{A15.1}$$
$$F_{frict}(\omega = 2\nu, A = 1) = \gamma\nu + J_0(1) - 2J_1(1) \approx -0.095. \tag{A15.2}$$
$$F_{frict}(\omega = 2\nu, A = 2) = \gamma\nu + 2(J_0(1) - 2J_1(1)) \approx -0.21. \tag{A15.3}$$

At $\gamma = 0.15$

$$F_{frict}(\omega = 2v, A = 0.5) = \gamma v + A(J_0(1) - 2J_1(1)) \approx 0.015 - 0.5 * 0.115 = -0.042. \tag{A15.4}$$

$$F_{frict}(\omega = 2v, A = 1) = \gamma v + J_0(1) - 2J_1(1) \approx -0.1. \tag{A15.5}$$

$$F_{frict}(\omega = 2v, A = 2) = \gamma v + 2(J_0(1) - 2J_1(1)) \approx -0.215. \tag{A15.6}$$

When perturbation occurs at high frequency $\omega = nv, n \gg 1$,

$$F_{frict} = \gamma v + \frac{vA}{4\pi}\int_0^{2\pi/v}\left(\cos(nvt + \alpha_1\sin(vt)) - \cos(nvt - \alpha_1\sin(vt))\right)dt \approx \gamma v. \tag{A16}$$

The expression Eq. (A1) could be used for qualitative description of friction force frequency dependence provided that spectral component with frequency $v$ is considerable. This occurs in mode-locked oscillation regime where allowed frequencies are multiples of $v$.

## APPENDIX B: FRICTION FORCE APPROXIMATION BY MODES OF DRIVING FREQUENCY, ITS SECOND AND THIRD HARMONICS

In the approximation
$$x = x_0 + vt + \alpha_1\sin(vt + \varphi_1) + \alpha_2\sin(2vt + \varphi_2) \tag{B1}$$

average input of periodic lattice potential is defined by following equation:

$$\begin{aligned}\frac{v}{2\pi}\int_0^{T=2\pi/v}\sin x \, dt =\ & -J_1(\alpha_1)(J_0(\alpha_2)\sin(x_0 - \varphi_1) + J_1(\alpha_2)\sin(x_0 + \varphi_1 - \varphi_2)) + \\ & +\sin vt(J_0(\alpha_1)J_0(\alpha_2)\cos x_0 + J_0(\alpha_1)J_1(\alpha_2)\cos(x_0 - \varphi_2)) + \\ & +\cos vt(J_0(\alpha_1)J_0(\alpha_2)\sin x_0 - J_0(\alpha_1)J_1(\alpha_2)\sin(x_0 - \varphi_2)) + \\ & +\sin 2vt(-2J_1(\alpha_1)J_1(\alpha_2)\cos(x_0 - \varphi_1)\cos\varphi_2 + J_1(\alpha_1)J_0(\alpha_2)\cos(x_0 + \varphi_1)) + \\ & +\cos 2vt(-2J_1(\alpha_1)J_1(\alpha_2)\cos(x_0 - \varphi_1)\sin\varphi_2 + J_1(\alpha_1)J_0(\alpha_2)\sin(x_0 + \varphi_1)) + \\ & +\sin 3vt J_0(\alpha_1)J_1(\alpha_2)\cos(x_0 + \varphi_2) + \cos 3vt J_0(\alpha_1)J_1(\alpha_2)\sin(x_0 + \varphi_2) - \\ & -\sin 4vt J_1(\alpha_1)J_1(\alpha_2)\cos(x_0 + \varphi_1 + \varphi_2) - \cos 4vt J_1(\alpha_1)J_1(\alpha_2)\sin(x_0 + \varphi_1 + \varphi_2).\end{aligned} \tag{B2}$$

In absence of perturbation ($A = 0$) collecting components with equal frequencies we obtain the system of equations to determine phases and amplitudes of modes

Constant term:
$$x_0 + \gamma v = J_1(\alpha_1)(J_0(\alpha_2)\sin(x_0 - \varphi_1) + J_1(\alpha_2)\sin(x_0 + \varphi_1 - \varphi_2)).$$
$\sin vt$:
$$\alpha_1((1 - v^2)\cos\varphi_1 - \gamma v\sin\varphi_1) = -J_0(\alpha_1)(J_0(\alpha_2)\cos x_0 + J_1(\alpha_2)\cos(x_0 - \varphi_2)).$$
$\cos 2vt$:
$$\alpha_1((1 - v^2)\sin\varphi_1 + \gamma v\cos\varphi_1) = -J_0(\alpha_1)(J_0(\alpha_2)\sin x_0 - J_1(\alpha_2)\sin(x_0 - \varphi_2)).$$
$\sin 2vt$:
$$\alpha_2((1 - 4v^2)\cos\varphi_2 - 2\gamma v\sin\varphi_2) = -J_1(\alpha_1)(J_0(\alpha_2)\cos(x_0 + \varphi_1) - 2J_1(\alpha_2)\cos(x_0 - \varphi_1)\cos\varphi_2).$$
$\cos 2vt$:
$$\alpha_2((1 - 4v^2)\sin\varphi_2 + 2\gamma v\cos\varphi_2) = -J_1(\alpha_1)(J_0(\alpha_2)\sin(x_0 + \varphi_1) - 2J_1(\alpha_2)\cos(x_0 - \varphi_1)\sin\varphi_2).$$

Friction force in absence of perturbation at $\omega_0 = 1$ is given by the following formula

$$F_{frict}(A = 0) = \gamma v - J_1(\alpha_1)\big(J_0(\alpha_2)\sin(x_0 - \varphi_1) + J_1(\alpha_2)\sin(x_0 + \varphi_1 - \varphi_2)\big). \quad \text{(B3)}$$

Input into average friction force for perturbation with frequency $\omega = v$ is

$$\frac{v}{2\pi}\int_0^{T=2\pi/v} A\sin vt \sin x =$$
$$= 0.5AJ_0(\alpha_1)\big(J_0(\alpha_2)\cos x_0 + J_1(\alpha_2)\cos(x_0 - \varphi_2)\big) +$$
$$-0.5AJ_1(\alpha_1)\sin vt\big(2J_0(\alpha_2)\sin x_0\cos\varphi_1 + J_1(\alpha_2)(\sin(x_0 + \varphi_1 - \varphi_2) - 2\cos(x_0 - \varphi_1)\sin\varphi_2)\big)$$
$$+0.5AJ_1(\alpha_1)\cos vt\big(J_0(\alpha_2)\cos(x_0 + \varphi_1) - 2\cos(x_0 - \varphi_1)\cos\varphi_2\big) +$$
$$+0.5A\sin 2vt J_0(\alpha_1)\sin x_0\big(J_0(\alpha_2) - 2J_1(\alpha_2)\cos\varphi_1\big)$$
$$-0.5A\cos 2vt J_0(\alpha_1)\big(J_0(\alpha_2)\cos x_0 + 2J_1(\alpha_2)\sin x_0\cos\varphi_1\big).$$
$$\quad \text{(B4)}$$

Input into average friction force for perturbation with frequency $\omega = 2v$ is determined as

$$\frac{v}{2\pi}\int_0^{T=2\pi/v} A\sin 2vt \sin x = 0.5A[$$
$$J_1(\alpha_1)\big(J_0(\alpha_2)\cos(x_0 + \varphi_1) - 2J_1(\alpha_2)\cos(x_0 - \varphi_1)\cos\varphi_2\big) +$$
$$+\sin vt J_0(\alpha_1)\sin x_0\big(J_0(\alpha_2) - J_1(\alpha_2)\cos\varphi_2\big) +$$
$$+\cos vt J_0(\alpha_1)\cos x_0\big(J_0(\alpha_2) + J_1(\alpha_2)\cos\varphi_2\big) +$$
$$+\sin 2vt J_1(\alpha_1)\big(-J_0(\alpha_2)\sin(x_0 - \varphi_1) + 2J_1(\alpha_2)\cos(x_0 + \varphi_1)\sin\varphi_2\big) -$$
$$-\cos 2vt J_1(\alpha_1)J_1(\alpha_2)\cos(x_0 + \varphi_1 + \varphi_2) +$$
$$+\sin 3vt J_0(\alpha_1)\big(J_0(\alpha_2)\sin x_0 - J_1(\alpha_2)\sin(x_0 - \varphi_2)\big) -$$
$$-\cos 3vt J_0(\alpha_1)\big(J_0(\alpha_2)\cos x_0 + J_1(\alpha_2)\cos(x_0 - \varphi_2)\big) +$$
$$+\sin 4vt J_1(\alpha_1)\big(J_0(\alpha_2)\sin(x_0 + \varphi_1) - J_1(\alpha_2)\cos(x_0 - \varphi_1)\sin\varphi_2\big) -$$
$$-\cos 4vt J_1(\alpha_1)\big(J_0(\alpha_2)\cos(x_0 + \varphi_1) - J_1(\alpha_2)\cos(x_0 - \varphi_1)\cos\varphi_2\big) +$$
$$+\sin 5vt J_0(\alpha_1)J_1(\alpha_2)\sin(x_0 + \varphi_2) - \cos 5vt J_0(\alpha_1)J_1(\alpha_2)\cos(x_0 + \varphi_2) -$$
$$-\sin 6vt J_1(\alpha_1)J_1(\alpha_2)\sin(x_0 + \varphi_1 + \varphi_2) + \cos 6vt J_1(\alpha_1)J_1(\alpha_2)\cos(x_0 + \varphi_1 + \varphi_2)].$$
$$\quad \text{(B5)}$$

When perturbation frequency resonant to driving parameter $v$ the friction force is following

$$F_{frict}(\omega = v) = \gamma v - J_1(\alpha_1)\big(J_0(\alpha_2)\sin(x_0 - \varphi_1) + J_1(\alpha_2)\sin(x_0 + \varphi_1 - \varphi_2)\big) +$$
$$+ 0.5AJ_0(\alpha_1)\big(J_0(\alpha_2)\cos x_0 + J_1(\alpha_2)\cos(x_0 - \varphi_2)\big). \quad \text{(B6)}$$

If frequency of perturbation is equal to second harmonic of driving parameter $v$ the friction force

$$F_{frict}(\omega = 2v) = \gamma v - J_1(\alpha_1)\big(J_0(\alpha_2)\sin(x_0 - \varphi_1) + J_1(\alpha_2)\sin(x_0 + \varphi_1 - \varphi_2)\big) +$$
$$+ 0.5AJ_1(\alpha_1)\big(J_0(\alpha_2)\cos(x_0 + \varphi_1) - 2J_1(\alpha_2)\cos(x_0 - \varphi_1)\cos\varphi_2\big). \quad \text{(B7)}$$

In three-mode approximation

$$x = x_0 + vt + \alpha_1\sin vt + \alpha_2\sin 2vt + \alpha_3\sin 3vt. \quad \text{(B8)}$$

Nonlinear term of lattice potential is represented with sum of frequency components

$$\begin{aligned}
\sin x = \sin(x_0 + vt &+ \alpha_1 \sin vt + \alpha_2 \sin 2vt + \alpha_3 \sin 3vt) = \\
= -\sin x_0 &\left( J_1(\alpha_1) J_0(\alpha_3)(J_0(\alpha_2) + J_1(\alpha_2)) + J_0(\alpha_1) J_1(\alpha_2) J_1(\alpha_3) \right) + \\
+\sin vt \cos x_0 &\left( J_0(\alpha_1) J_0(\alpha_3)(J_0(\alpha_2) + J_1(\alpha_2)) + J_1(\alpha_1) J_1(\alpha_3)(J_0(\alpha_2) - J_1(\alpha_2)) \right) + \\
+\cos vt \sin x_0 &\left( J_0(\alpha_1) J_0(\alpha_3) - J_1(\alpha_1) J_1(\alpha_3) \right)(J_0(\alpha_2) - J_1(\alpha_2)) + \\
+\sin 2vt \cos x_0 &\left( J_1(\alpha_1) J_0(\alpha_3)(J_0(\alpha_2) - 2J_1(\alpha_2)) + J_0(\alpha_1) J_1(\alpha_3)(J_0(\alpha_2) - J_1(\alpha_2)) \right) + \\
+\cos 2vt \sin x_0 &\left( J_1(\alpha_1) J_0(\alpha_2) J_0(\alpha_3) - J_0(\alpha_1) J_1(\alpha_3)(J_0(\alpha_2) + J_1(\alpha_2)) \right) + \\
+\sin 3vt \cos x_0 &\left( J_0(\alpha_1) J_1(\alpha_2) J_0(\alpha_3) - 2J_1(\alpha_1) J_1(\alpha_3)(J_0(\alpha_2) + J_1(\alpha_2)) \right) + \\
+\cos 3vt \sin x_0 &\, J_1(\alpha_1) J_0(\alpha_2) J_0(\alpha_3) + f(4vt, 5vt..)
\end{aligned}$$
(B9)

Input of lattice potential and perturbation and into average friction force is

$$\begin{aligned}
\frac{v}{2\pi} &\int_0^{2\pi/v} (1 + A\sin\omega t) \sin x \, dt = \\
-\sin x_0 &\left( J_1(\alpha_1) J_0(\alpha_3)(J_0(\alpha_2) + J_1(\alpha_2)) + J_0(\alpha_1) J_1(\alpha_2) J_1(\alpha_3) \right) + \\
+\frac{A}{2}\cos x_0 \cdot &\Big( \delta(\omega - v) \left( J_0(\alpha_1) J_0(\alpha_3)(J_0(\alpha_2) + J_1(\alpha_2)) + J_1(\alpha_1) J_1(\alpha_3)(J_0(\alpha_2) - J_1(\alpha_2)) \right) + \\
+\delta(\omega - 2v) &\left( J_1(\alpha_1) J_0(\alpha_3)(J_0(\alpha_2) - 2J_1(\alpha_2)) + J_0(\alpha_1) J_1(\alpha_3)(J_0(\alpha_2) - J_1(\alpha_2)) \right) + \\
+\delta(\omega - 3v) &\left( J_0(\alpha_1) J_1(\alpha_2) J_0(\alpha_3) - 2J_1(\alpha_1) J_1(\alpha_3)(J_0(\alpha_2) + J_1(\alpha_2)) \right) \Big)
\end{aligned}$$
(B10)

Average friction force in approximation of Eq. (B8) for perturbation with frequency $\omega$ is estimates by expression

$$\begin{aligned}
F_{frict}(\omega = v) = \gamma v &- \sin x_0 \left( J_1(\alpha_1) J_0(\alpha_3)(J_0(\alpha_2) + J_1(\alpha_2)) + J_0(\alpha_1) J_1(\alpha_2) J_1(\alpha_3) \right) + \\
+ \frac{A}{2}\cos x_0 &\left( J_0(\alpha_1) J_0(\alpha_3)(J_0(\alpha_2) + J_1(\alpha_2)) + J_1(\alpha_1) J_1(\alpha_3)(J_0(\alpha_2) - J_1(\alpha_2)) \right).
\end{aligned}$$
(B11)

Friction force for perturbation with frequency $2\omega$

$$\begin{aligned}
F_{frict}(\omega = 2v) = \gamma v &- \sin x_0 \left( J_1(\alpha_1) J_0(\alpha_3)(J_0(\alpha_2) + J_1(\alpha_2)) + J_0(\alpha_1) J_1(\alpha_2) J_1(\alpha_3) \right) + \\
+ \frac{A}{2}\cos x_0 &\left( J_1(\alpha_1) J_0(\alpha_3)(J_0(\alpha_2) - 2J_1(\alpha_2)) + J_0(\alpha_1) J_1(\alpha_3)(J_0(\alpha_2) - J_1(\alpha_2)) \right).
\end{aligned}$$
(B12)

Friction force for perturbation with frequency $3\omega$

$$\begin{aligned}
F_{frict}(\omega = 3v) = \gamma v &- \sin x_0 \left( J_1(\alpha_1) J_0(\alpha_3)(J_0(\alpha_2) + J_1(\alpha_2)) + J_0(\alpha_1) J_1(\alpha_2) J_1(\alpha_3) \right) + \\
+ \frac{A}{2}\cos x_0 &\left( J_0(\alpha_1) J_1(\alpha_2) J_0(\alpha_3) - 2J_1(\alpha_1) J_1(\alpha_3)(J_0(\alpha_2) + J_1(\alpha_2)) \right).
\end{aligned}$$
(B13)

## APPENDIX C: THREE-MODE APPROXIMATION OF FRICTION FORCE WITH CENTRAL, STOKES AND ANTI-STOKES FREQUENCIES

In this three-mode approximation of friction force in oscillator model it is assumed that perturbation frequency is high comparing to driving parameter $\omega \gg v$. Oscillator displacement is approximated as

$$x(t) = x_0 + vt + \alpha_{\omega-v}\sin(\omega - v)t + \alpha_\omega \sin\omega t + \alpha_{\omega+v}\sin(\omega + v)t. \quad (C1)$$

At $\omega = nv$, where $n$ - integer, averaging of Eq. (C1) gives zero input into friction force.

The model equation Eq. (5) contains nonlinear term $\sin x$ which would provide nonzero input into average friction force.

$$<\sin x> = <\sin(x_0 + vt + \alpha_{\omega-v}\sin(\omega-v)t + \alpha_\omega \sin\omega t + \alpha_{\omega+v}\sin(\omega+v)t)> = $$
$$= -\sin x_0 J_1(\alpha_\omega)\big(J_0(\alpha_{\omega-v})J_1(\alpha_{\omega+v}) + J_0(\alpha_{\omega+v})J_1(\alpha_{\omega-v})\big). \quad (C2)$$

$$<A\sin\omega t \sin x> = 1/2 A J_0(\alpha_\omega)\cos x_0 \big(J_0(\alpha_{\omega-v})J_1(\alpha_{\omega+v}) + J_0(\alpha_{\omega+v})J_1(\alpha_{\omega-v})\big). \quad (C.3)$$

Average friction force is defined as

$$F_{frict}(\omega) = \frac{1}{T}\int_0^T (\ddot{x} + \gamma\dot{x} + (1 + A\sin\omega t)\sin x)dt = $$
$$= \gamma v + \left(-\sin x_0 J_1(\alpha_\omega) + \frac{A}{2}J_0(\alpha_\omega)\cos x_0\right)\big(J_0(\alpha_{\omega-v})J_1(\alpha_{\omega+v}) + J_0(\alpha_{\omega+v})J_1(\alpha_{\omega-v})\big), \quad (C4)$$

where the coefficients $\sin x_0$ and $\alpha_{\omega\pm v}$ are defined as

$$\sin x_0 = -\frac{\alpha_1 \gamma v}{(J_0(\alpha_1)J_0(\alpha_3) - J_1(\alpha_1)J_1(\alpha_3))(J_0(\alpha_2) - J_1(\alpha_2))}. \quad (C5)$$

$$\alpha_{\omega\pm v}(1 - (\omega \pm v)^2) + J_0(\alpha_{\omega-v})J_1(\alpha_\omega)J_0(\alpha_{\omega+v})\cos x_0 + \frac{A}{2}\sin x_0 J_0(\alpha_{\omega-v})J_0(\alpha_\omega)J_0(\alpha_{\omega+v}) = 0, \quad (C6.1)$$

$$\gamma(\omega \pm v)\alpha_{\omega\pm v} \pm J_0(\alpha_{\omega-v})J_1(\alpha_\omega)J_0(\alpha_{\omega+v})\sin x_0 \mp \frac{A}{2}\cos x_0 J_0(\alpha_{\omega-v})J_0(\alpha_\omega)J_0(\alpha_{\omega+v}) = 0. \quad (C6.2)$$

In the simplified approximation without spectral component at central frequency $\omega$ of perturbation the oscillator coordinate is approximated as

$$x = x_0 + vt + \alpha_{\omega-v}\sin(\omega - v)t + \alpha_{\omega+v}\sin(\omega + v)t, \quad (C7)$$

$$\dot{x} = v + (\omega + v)\alpha_{\omega+v}\cos(\omega + v)t + (\omega - v)\alpha_{\omega-v}\cos(\omega - v)t,$$

$$\ddot{x} = -\alpha_{\omega+v}(\omega + v)^2\sin(\omega + v)t - (\omega - v)^2 \alpha_{\omega-v}\sin(\omega - v)t.$$

$$\sin x = \sin(x_0 + vt + \alpha_{\omega+v}\sin(\omega + v)t + \alpha_{\omega-v}\sin(\omega - v)t) =$$
$$\sin(x_0 + vt)\cos(\alpha_{\omega+v}\sin(\omega + v)t + \alpha_{\omega-v}\sin(\omega - v)t) +$$
$$\cos(x_0 + vt)\sin(\alpha_{\omega+v}\sin(\omega + v)t + \alpha_{\omega-v}\sin(\omega - v)t) =$$
$$\sin(x_0 + vt)(J_0(\alpha_{\omega+v})J_0(\alpha_{\omega-v}) - 4J_1(\alpha_{\omega+v})J_1(\alpha_{\omega-v})\sin(\omega + v)t\sin(\omega - v)t)$$
$$+2\cos(x_0 + vt)(J_0(\alpha_{\omega-v})J_1(\alpha_{\omega+v})\sin(\omega + v)t + J_0(\alpha_{\omega+v})J_1(\alpha_{\omega-v})\sin(\omega - v)t) =$$
$$J_0(\alpha_{\omega+v})J_0(\alpha_{\omega-v})\sin(x_0 + vt) -$$
$$-J_1(\alpha_{\omega+v})J_1(\alpha_{\omega-v})\Big(\sin(3vt + x_0) - \sin(vt - x_0) - \sin((2\omega + v)t + x_0) + \sin((2\omega - v)t - x_0)\Big) +$$
$$+J_0(\alpha_{\omega-v})J_1(\alpha_{\omega+v})\Big(\sin((\omega + 2v)t + x_0) + \sin(\omega t - x_0)\Big) +$$
$$+J_0(\alpha_{\omega+v})J_1(\alpha_{\omega-v})\Big(\sin((\omega - 2v)t - x_0) + \sin(\omega t + x_0)\Big).$$
(C8)

$$A\sin\omega t\sin x = \frac{A}{2}J_0(\alpha_{\omega+v})J_0(\alpha_{\omega-v})\Big(\cos((\omega - v)t - x_0) - \cos((\omega + v)t + x_0)\Big) -$$
$$-\frac{A}{2}J_1(\alpha_{\omega+v})J_1(\alpha_{\omega-v})f(\omega \pm 3v, \omega \pm v, 3\omega \pm v) +$$
$$+\frac{A}{2}J_0(\alpha_{\omega-v})J_1(\alpha_{\omega+v})\big(\cos(2vt + x_0) - \cos(2(\omega + v)t + x_0) + \cos x_0 - \cos(2\omega t - x_0)\big) +$$
$$+\frac{A}{2}J_0(\alpha_{\omega+v})J_1(\alpha_{\omega-v})\big(\cos(2vt + x_0) - \cos(2(\omega - v)t - x_0) + \cos x_0 - \cos(2\omega t + x_0)\big),$$
(C9)

where
$$f(\omega \pm 3v, \omega \pm v, 3\omega \pm v) =$$
$$= \cos((\omega - 3v)t - x_0) - \cos((\omega + 3v)t + x_0) - \cos((\omega - v)t + x_0)$$
$$+ \cos((\omega - v)t - x_0) - \cos((\omega + v)t - x_0) - \cos((\omega + v)t + x_0)$$
$$+ \cos((3\omega + v)t + x_0) - \cos((3\omega - v)t - x_0).$$

Substitution of Eqs. (C7)-(C9) into Eq. (2) and averaging gives the expression of average friction force

$$F_{frict} = \frac{1}{T}\int_0^T (\ddot{x} + \gamma\dot{x} + (1 + A\sin\omega t)\sin x)dt =$$
$$= \gamma v + \frac{1}{2}A\cos x_0\big(J_0(\alpha_{\omega-v})J_1(\alpha_{\omega+v}) + J_0(\alpha_{\omega+v})J_1(\alpha_{\omega-v})\big). \quad (C10)$$

In this derivation it is supposed that $\omega$ is a multiple of $v$, i.e. $\omega = nv$, $n \gg 1$. To calculate $\cos x_0$ and the coefficients $\alpha_{\omega\pm v}$ the terms of each frequency component $\omega \pm v$ are selected from equation

$$\left(\alpha_{\omega\pm v}(1 - (\omega \pm v)^2) + \frac{A}{2}\sin x_0\big(J_0(\alpha_{\omega+v})J_0(\alpha_{\omega-v}) - 2J_1(\alpha_{\omega+v})J_1(\alpha_{\omega-v})\big)\right)\sin(\omega \pm v)t +$$
$$+ \left(\gamma(\omega \pm v)\alpha_{\omega\pm v} \mp \frac{A}{2}J_0(\alpha_{\omega+v})J_0(\alpha_{\omega-v})\cos x_0\right)\cos(\omega \pm v)t = 0.$$
(C11)

The expressions to define $\cos x_0$ and coefficients $\alpha_{\omega\pm v}$ are

$$\cos x_0 = \pm\frac{2\gamma(\omega\pm v)\alpha_{\omega\pm v}}{AJ_0(\alpha_{\omega+v})J_0(\alpha_{\omega-v})}, \quad (C12)$$

$$\alpha_{\omega\pm v} = -\frac{A\sin x_0\big(J_0(\alpha_{\omega+v})J_0(\alpha_{\omega-v}) - 2J_1(\alpha_{\omega+v})J_1(\alpha_{\omega-v})\big)}{2(1-(\omega\pm v)^2)}. \quad (C13)$$

Then $(\omega + v)\alpha_{\omega+v} = -(\omega - v)\alpha_{\omega-v}$, and the ratio of anti-Stokes and Stokes components is

$$\frac{\alpha_{\omega+v}}{\alpha_{\omega-v}} = -\frac{\omega-v}{\omega+v}. \tag{C14}$$

For perturbation at frequency $\omega > v$ the amplitude of Stokes frequency component is greater than the amplitude of anti-Stokes component $|\alpha_{\omega+v}| < |\alpha_{\omega-v}|$. Substituting the Eq. (C12) into Eq. (C10) results in formula for average friction force in approximation of Eq. (C7):

$$F_{frict} = \gamma \left( v \pm \frac{(\omega \pm v)\alpha_{\omega \pm v}(J_1(\alpha_{\omega+v})J_0(\alpha_{\omega-v}) + J_1(\alpha_{\omega-v})J_0(\alpha_{\omega+v}))}{J_0(\alpha_{\omega+v})J_0(\alpha_{\omega-v})} \right). \tag{C15}$$

From this expression it is follows that the friction force in the oscillator model could be negative.